\documentclass[reprint,superscriptaddress,amsmath,amssymb,aps,pra,floatfix]{revtex4-1}

\usepackage{graphicx}
\usepackage{dcolumn}
\usepackage{bm}
\usepackage{hyperref}
\usepackage{commath}
\usepackage{siunitx}
\usepackage{subfigure}
\usepackage{placeins}

\usepackage{color}
\usepackage{upgreek}

\newcommand{\av}[1]{\langle #1 \rangle}
\newcommand{\ct}{\cos \theta}

\newcommand{\muT}{\upmu {\rm T}}

\renewcommand{\arraystretch}{1.5}
\newcommand{\R}{\mathcal{R}}
\newcommand{\sphi}{\sin \phi}
\newcommand{\cphi}{\cos \phi}
\newcommand{\sph}[1]{\sin #1 \phi}
\newcommand{\cph}[1]{\cos #1 \phi}
\newcommand{\rp}[1]{\rho^#1}
\newcommand{\zp}[1]{z^#1}
\newcommand{\Gsq}[2]{(G_{#1,-#2}^2 + G_{#1,#2}^2)}

\newcommand{\guil}[1]{``#1"}
\newcommand{\bt}{\langle B_{\rm T}^2 \rangle}
\newcommand{\G}{\widehat{G}}

\begin{document}

\preprint{APS/123-QED}

\title{Mapping of the magnetic field to correct systematic effects in a neutron electric dipole moment experiment}

\author{C.~Abel}
\affiliation{Department of Physics and Astronomy, University of Sussex, Falmer, Brighton BN1 9QH, UK}
\author{N.\,J.~Ayres}
\email[Corresponding author: ]{ayresn@phys.ethz.ch}
\affiliation{Department of Physics and Astronomy, University of Sussex, Falmer, Brighton BN1 9QH, UK}
\affiliation{ETH Z\"{u}rich, Institute for Particle Physics and Astrophysics, CH-8093 Z\"{u}rich, Switzerland}
\author{G.~Ban}
\affiliation{LPC Caen, ENSICAEN, Universit\'{e} de Caen, CNRS/IN2P3, 14000 Caen, France}
\author{G.~Bison}
\affiliation{Paul Scherrer Institut, CH-5232 Villigen PSI, Switzerland}
\author{K.~Bodek}
\affiliation{Marian Smoluchowski Institute of Physics, Jagiellonian University, 30-348 Cracow, Poland}
\author{V.~Bondar}
\affiliation{ETH Z\"{u}rich, Institute for Particle Physics and Astrophysics, CH-8093 Z\"{u}rich, Switzerland}
\affiliation{Paul Scherrer Institut, CH-5232 Villigen PSI,
Switzerland}
\affiliation{Instituut voor Kern- en Stralingsfysica, University of Leuven, B-3001 Leuven, Belgium}
\author{E.~Chanel}
\affiliation{Laboratory for High Energy Physics and Albert Einstein Center for Fundamental Physics, University of Bern, CH-3012 Bern, Switzerland}
\author{P.-J.~Chiu}
\affiliation{ETH Z\"{u}rich, Institute for Particle Physics and Astrophysics, CH-8093 Z\"{u}rich, Switzerland}
\affiliation{Paul Scherrer Institut, CH-5232 Villigen PSI, Switzerland}
\author{B.~Cl\'ement}
\affiliation{Universit\'e Grenoble Alpes, CNRS, Grenoble INP, LPSC-IN2P3, 38026 Grenoble, France}
\author{C.\,B.~Crawford} 
\affiliation{University of Kentucky, Lexington, KY 40506, USA}
\author{M.~Daum}
\affiliation{Paul Scherrer Institut, CH-5232 Villigen PSI, Switzerland}
\author{S.~Emmenegger}
\affiliation{ETH Z\"{u}rich, Institute for Particle Physics and Astrophysics, CH-8093 Z\"{u}rich, Switzerland}
\author{L.~Ferraris-Bouchez}
\email[Corresponding author: ]{ferraris@lpsc.in2p3.fr}
\affiliation{Universit\'e Grenoble Alpes, CNRS, Grenoble INP, LPSC-IN2P3, 38026 Grenoble, France}
\author{M.~Fertl}
\affiliation{Institut f\"{u}r Physik, Johannes-Gutenberg-Universit\"{a}t, D-55128 Mainz, Germany}
\author{P.~Flaux}
\affiliation{LPC Caen, ENSICAEN, Universit\'{e} de Caen, CNRS/IN2P3, 14000 Caen, France}
\author{A.~Fratangelo}
\affiliation{Laboratory for High Energy Physics and Albert Einstein Center for Fundamental Physics, University of Bern, CH-3012 Bern, Switzerland}
\author{W.\,C.~Griffith}
\affiliation{Department of Physics and Astronomy, University of Sussex, Falmer, Brighton BN1 9QH, UK}
\author{Z.\,D.~Gruji\'c}
\affiliation{Physics Department, University of Fribourg, CH-1700 Fribourg, Switzerland}
\affiliation{Institute of Physics Belgrade, University of Belgrade, 11080 Belgrade, Serbia}
\author{P.\,G.~Harris}
\affiliation{Department of Physics and Astronomy, University of Sussex, Falmer, Brighton BN1 9QH, UK}
\author{L.~Hayen}
\altaffiliation[Present address: ]{Department of Physics, North Carolina State University, Raleigh, NC 27695, USA}
\affiliation{Instituut voor Kern- en Stralingsfysica, University of Leuven, B-3001 Leuven, Belgium}
\author{N.~Hild}
\affiliation{ETH Z\"{u}rich, Institute for Particle Physics and Astrophysics, CH-8093 Z\"{u}rich, Switzerland}
\affiliation{Paul Scherrer Institut, CH-5232 Villigen PSI, Switzerland}
\author{M.~Kasprzak}
\affiliation{Paul Scherrer Institut, CH-5232 Villigen PSI, Switzerland}
\affiliation{Instituut voor Kern- en Stralingsfysica, University of Leuven, B-3001 Leuven, Belgium}
\affiliation{Physics Department, University of Fribourg, CH-1700 Fribourg, Switzerland}
\author{K.~Kirch}
\affiliation{ETH Z\"{u}rich, Institute for Particle Physics and Astrophysics, CH-8093 Z\"{u}rich, Switzerland}
\affiliation{Paul Scherrer Institut, CH-5232 Villigen PSI,
Switzerland}
\author{P.~Knowles}
\affiliation{Physics Department, University of Fribourg, CH-1700 Fribourg, Switzerland}
\author{H.-C.~Koch}
\affiliation{Paul Scherrer Institut, CH-5232 Villigen PSI, Switzerland}
\affiliation{Institut f\"{u}r Physik, Johannes-Gutenberg-Universit\"{a}t, D-55128 Mainz, Germany}
\affiliation{Physics Department, University of Fribourg, CH-1700 Fribourg, Switzerland}
%
%
\author{P.A.~Koss}
\altaffiliation[Present address: ]{Fraunhofer-Institut f\"{u}r Physikalische Messtechnik IPM, 79110 Freiburg i. Breisgau, Germany}
\affiliation{Instituut voor Kern- en Stralingsfysica, University of Leuven, B-3001 Leuven, Belgium}
\author{A.~Kozela}
\affiliation{Henryk Niedwodniczanski Institute for Nuclear Physics, 31-342 Cracow, Poland}
\author{J.~Krempel}
\affiliation{ETH Z\"{u}rich, Institute for Particle Physics and Astrophysics, CH-8093 Z\"{u}rich, Switzerland}
\author{B.~Lauss}
\affiliation{Paul Scherrer Institut, CH-5232 Villigen PSI,
Switzerland}
\author{T.~Lefort}
\affiliation{LPC Caen, ENSICAEN, Universit\'{e} de Caen, CNRS/IN2P3, 14000 Caen, France}
\author{Y.~Lemi\`{e}re}
\affiliation{LPC Caen, ENSICAEN, Universit\'{e} de Caen, CNRS/IN2P3, 14000 Caen, France}
\author{P.~Mohanmurthy}
\affiliation{ETH Z\"{u}rich, Institute for Particle Physics and Astrophysics, CH-8093 Z\"{u}rich, Switzerland}
\affiliation{Paul Scherrer Institut, CH-5232 Villigen PSI,
Switzerland}
\altaffiliation[Present address: ]{University of Chicago, Chicago, IL 60637, USA}
\author{O.~Naviliat-Cuncic} 
\affiliation{LPC Caen, ENSICAEN, Universit\'{e} de Caen, CNRS/IN2P3, 14000 Caen, France}
\author{D.~Pais}
\affiliation{ETH Z\"{u}rich, Institute for Particle Physics and Astrophysics, CH-8093 Z\"{u}rich, Switzerland}
\affiliation{Paul Scherrer Institut, CH-5232 Villigen PSI,
Switzerland}
\author{F.M.~Piegsa}
\affiliation{Laboratory for High Energy Physics and Albert Einstein Center for Fundamental Physics, University of Bern, CH-3012 Bern, Switzerland}
\author{G.~Pignol}
\affiliation{Universit\'e Grenoble Alpes, CNRS, Grenoble INP, LPSC-IN2P3, 38026 Grenoble, France}
\author{P.\,N.~Prashanth}
\affiliation{Paul Scherrer Institut, CH-5232 Villigen PSI, Switzerland}
\affiliation{Instituut voor Kern- en Stralingsfysica, University of Leuven, B-3001 Leuven, Belgium}
\author{G.~Qu\'{e}m\'{e}ner}
\affiliation{LPC Caen, ENSICAEN, Universit\'{e} de Caen, CNRS/IN2P3, 14000 Caen, France}
\author{M.~Rawlik}
\altaffiliation[Present address:  ]{Paul Scherrer Institut, CH-5232 Villigen PSI, Switzerland}
\affiliation{ETH Z\"{u}rich, Institute for Particle Physics and Astrophysics, CH-8093 Z\"{u}rich, Switzerland}
\author{D.~Ries}
\affiliation{Department of Chemistry - TRIGA site, Johannes Gutenberg University
Mainz, D-55128 Mainz, Germany}
%
\author{D.~Rebreyend}
\affiliation{Universit\'e Grenoble Alpes, CNRS, Grenoble INP, LPSC-IN2P3, 38026 Grenoble, France}
\author{S.~Roccia}
\affiliation{Universit\'e Grenoble Alpes, CNRS, Grenoble INP, LPSC-IN2P3, 38026 Grenoble, France}
\affiliation{Institut Laue-Langevin, CS 20156 F-38042 Grenoble Cedex 9, France}
\author{D.~Rozpedzik}
\affiliation{Marian Smoluchowski Institute of Physics, Jagiellonian University, 30-348 Cracow, Poland}
\author{P.~Schmidt-Wellenburg}
\email[Corresponding author: ]{philipp.schmidt-wellenburg@psi.ch} \affiliation{Paul Scherrer Institut, CH-5232 Villigen PSI, Switzerland}
\author{A.~Schnabel}
\affiliation{Physikalisch Technische Bundesanstalt, D-10587 Berlin, Germany}
\author{N.~Severijns}
\affiliation{Instituut voor Kern- en Stralingsfysica, University of Leuven, B-3001 Leuven, Belgium}
\author{J.\,A.~Thorne}
\affiliation{Department of Physics and Astronomy, University of Sussex, Falmer, Brighton BN1 9QH, UK}
\affiliation{Laboratory for High Energy Physics and Albert Einstein Center for Fundamental Physics, University of Bern, CH-3012 Bern, Switzerland}
\author{R.~Virot}
\affiliation{Universit\'e Grenoble Alpes, CNRS, Grenoble INP, LPSC-IN2P3, 38026 Grenoble, France}
\author{A.~Weis}
\affiliation{Physics Department, University of Fribourg, CH-1700 Fribourg, Switzerland}
\author{E.~Wursten}
\altaffiliation[Present address: ]{CERN, 1211 Gen\`eve, Switzerland}
\affiliation{Instituut voor Kern- en Stralingsfysica, University of Leuven, B-3001 Leuven, Belgium}
\author{G.~Wyszynski}
\affiliation{ETH Z\"{u}rich, Institute for Particle Physics and Astrophysics, CH-8093 Z\"{u}rich, Switzerland}
\affiliation{Marian Smoluchowski Institute of Physics, Jagiellonian University, 30-348 Cracow, Poland}
\author{J.~Zejma}
\affiliation{Marian Smoluchowski Institute of Physics, Jagiellonian University, 30-348 Cracow, Poland}
\author{G.~Zsigmond}
\affiliation{Paul Scherrer Institut, CH-5232 Villigen PSI, Switzerland}
%


\date{\today}

\begin{abstract}
Experiments dedicated to the measurement of the electric dipole moment of the neutron require outstanding control of the magnetic field uniformity. The neutron electric dipole moment (nEDM) experiment at the Paul Scherrer Institute 
uses a $^{199}{\rm Hg}$ co-magnetometer to precisely monitor temporal magnetic field variations. This co-magnetometer, in the presence of field non-uniformity, is however responsible for the largest systematic effect of this measurement. To evaluate and correct that effect, offline measurements of the field non-uniformity were performed during mapping campaigns in 2013, 2014 and 2017. We present the results of these campaigns, and the improvement the correction of this effect brings to the neutron electric dipole moment measurement.
\end{abstract}

\pacs{Valid PACS appear here}%
\maketitle

\section{\label{sec:Intro}Introduction}
Discovering a non-zero electric dipole moment (EDM) of a simple spin-1/2 particle, like the neutron, would have far-reaching implications. 
Indeed, the existence of such a moment implies a violation of time-reversal invariance T, and therefore a violation of CP symmetry, under the assumption that combined CPT symmetry holds~\cite{Luders1954}. 
The electroweak theory of the Standard Model of particle physics predicts tiny values for all subatomic particles' EDMs, making them background free observables and ideal probes of new physics beyond the Standard Model. 
The experimental search for the neutron EDM has been an important research topic since the early 1950s~\cite{purcell1950}. 
There has been an improvement of six orders of magnitude in the measurement precision between the first experiment~\cite{Smith1957} with a beam of neutrons and the most recent measurement~\cite{Abel2020_2} performed at the ultracold neutron (UCN) source~\cite{Bison2020} of the Paul Scherrer Institute (PSI) by the nEDM collaboration.
However, the measured neutron EDM is still compatible with zero:
\begin{equation}
d_\mathrm{n} = ( 0.0 \pm 1.1_{\rm stat} \pm 0.2_{\rm sys} ) \times 10^{-26} \, e \, {\rm cm}.
\end{equation}
This result was obtained with a substantially refitted apparatus originally developed by the Sussex/RAL/ILL collaboration~\cite{Baker2014}, which had given the previous most stringent limit \cite{Pendlebury2015} when running at the Institut Laue-Langevin (ILL). 
It was moved to PSI in 2009, and was then comprehensively upgraded and operated for several years, until autumn 2017. 
As with almost all other contemporary or future nEDM projects, the PSI nEDM experiment used ultracold neutrons (UCN) stored in a bottle for hundreds of seconds. The bottle was a cylindrical chamber of height $H=\SI{12}{cm}$ and radius $R =\SI{23.5}{cm}$. 
It sat coaxially in a stable and uniform vertical magnetic field with a magnitude of $B_0 \approx \SI{1}{\muT}$ in which the neutrons' spins precessed at the Larmor frequency of nominally $f_\mathrm{n}\approx\SI{30}{Hz}$. 
An electric field $E$ of \SI{11}{kV/cm} was also applied, either parallel or anti-parallel to the magnetic field.

The experimental method deployed to search for an nEDM is a precise measurement of the Larmor precession frequency, $f_\mathrm{n}$, of the neutrons' spins in the chamber with the Ramsey technique of (time)-separated oscillatory fields~\cite{Ramsey1950}. 
The EDM can then be extracted from the difference of frequencies between parallel and anti-parallel fields, $d_n = \pi \hbar (f_{n, \uparrow \downarrow} - f_{n, \uparrow \uparrow}) / 2E$.
In these experiments, the control of the magnetic field is the most important experimental challenge. 
Time fluctuations of $B_0$ must be monitored in real-time. 
For this reason, in the experiment~\cite{Abel2020_2,Pendlebury2015, Baker2014}, spin-polarized $^{199}$Hg atoms filled the precession chamber with the neutrons and were used as a co-magnetometer. 
The drifts of the magnetic field were corrected using the time-averaged precession frequency of the mercury atoms' spins $f_{\rm Hg}\approx \SI{7.6}{Hz}$ through the relation $f_{\rm Hg} = \gamma_{\rm Hg} B_0/(2\pi)$, where $\gamma_{\rm Hg}$ is the mercury gyromagnetic ratio. 
To maintain neutron spin coherence over the Ramsey cycle, a field uniformity better than \SI{1}{nT} must be achieved inside the chamber~\cite{Abel2019}.

This article is the third episode of a trilogy of papers dedicated to statistical and systematic uncertainties in nEDM searches due to the non-uniformity (gradients) of the magnetic field. 
The first article~\cite{Abel2019} describes the effects of magnetic-field non-uniformity for nEDM experiments.
Field inhomogeneities accelerate the depolarization of the neutrons, causing a loss of statistical sensitivity.
Simultaneously, they also cause systematic shifts in the neutron or mercury spin precession frequency.
The second paper explains how we limit the sensitivity loss in the PSI experiment. 
This is achieved using an \textit{in situ} magnetic-field homogenization strategy using an array of 16 Cs magnetometers~\cite{Abel2020}. 
However, the uniformity achieved thanks to this method was not enough to keep the systematic effects sufficiently low. 
We had then to characterize the magnetic field non-uniformity in order to correct for these effects.
In this article, we present this characterization: an offline mapping of the magnetic field. 
First, we will summarize the systematic effects induced by the non-uniformity that need to be evaluated. 
Then, we will describe the experiment's magnetic field and the mapping measurements. 
Finally, we will detail the mapping analysis and present its results.

\section{Systematic effects related to field non-uniformity\label{Sec:SystematicEffects}}
Critical for the extraction of the nEDM from the difference of precession frequencies $f_\mathrm{n}$ of the stored neutrons exposed to a positive and negative electric field is the control for coincidental or correlated changes in the magnetic field $B$. 
For this purpose $B$ is monitored using the  $^{199}$Hg co-magnetometer.
%
The largest systematic effect in this measurement, the so-called false EDM effect, arises from the combination of motional magnetic fields from the relativistic transformation of the large electric field into the rest frame of the thermal mercury atoms, which in the presence of a non-uniform magnetic field causes a shift in precession frequency linear in $E$, the same signature as a real electric dipole moment.
Other frequency shifts not linked to $E$ do not directly cause a systematic effect, they can indirectly interfere with the correction of the effect, and thus contribute to an overall systematic.
A full overview of all relevant systematic effects can be found in Table~I of Ref.~\cite{Abel2020_2}.


The primary purpose of the offline field mapping measurement detailed in this article is to measure the magnetic field non-uniformity over the precession-chamber volume.
As explained in~\cite{Abel2019}, we use a harmonic polynomial expansion to describe the field. 
In cylindrical coordinates ($\rho, \phi, z$), this expansion can be written as follows:
\begin{equation}
\label{eq:HarmonicPolynomialExpansion}
\vec{B}(\vec{r}) = \sum_{l,m} G_{l,m}
\left[
\begin{aligned}
&\Pi_{\rho,l,m}(\vec{r}) \\
&\Pi_{\phi,l,m}(\vec{r}) \\
&\Pi_{z,l,m}(\vec{r})
\end{aligned}
\right],
\end{equation} 
where the functions $\vec{\Pi}_{l,m}$ are products of a polynomial of order $l$ in $\rho, z$ and a trigonometric function in $m \phi$, and $G_{l,m}$ are the expansion coefficients, which will be called gradients in the rest of this article. 
Expressions for the first eighty modes in cylindrical coordinates, all modes $l\leq 7$, can be found in Tables~\ref{adequateCylindrical} to~\ref{adequateCylindrical3} in Appendix~\ref{appendix_polyCynlindrical}. Note that at each ``order'' $l$, polynomials with $-l-1 \leq m \leq +l+1$ exist.

Section~\ref{sec:hgIndFalseNEDM} discusses a frequency shift linear in $E$ which mimics the signature of an electric dipole moment signal.
Section~\ref{sec:trans} describes an effect independent of $E$, but which must be controlled to enable our correction strategy.
In Section~\ref{sec:CorrStrat} we describe an effect caused by vertical magnetic-field gradients, independent of $E$, but which inverts with the sign of $B$, which we make use of to elegantly eliminate the first effect described.
Finally, a significant shift in the measured frequencies caused by measuring in a rotating reference frame on Earth, not related to the magnetic field homogeneity, but reversing in sign with $B$ and therefore relevant to our correction strategy, is described in Appendix \ref{sec:rotation}.


\subsection{Mercury-induced false neutron EDM\label{sec:hgIndFalseNEDM}}
The dominant systematic effect in the measurement of the neutron EDM at PSI was the motional false EDM\@.
It is caused by the combination of non-uniformity of the magnetic field and a relativistic motional field experienced by the particles. 
It induces a linear-in-electric-field frequency shift, which is exactly the kind of signal a true neutron EDM would produce.
This shift has been extensively studied theoretically~\cite{Pignol2015b,Pendlebury2004,Lamoreaux2005,Barabanov2006,Clayton2011,Pignol2012,Swank2012,Steyerl2014,Golub2015,Swank2016} and discussed more specifically for the nEDM experiment at PSI in~\cite{Abel2019}. 
It can be split in two components: a direct effect due to the neutron and an indirect one from the mercury comagnetometer, which enters the neutron EDM measurement by contaminating the correction for magnetic field drifts. Nevertheless, use of the comagnetometer to control for random drifts in the magnetic field (uncorrelated with $E$) is required to achieve reasonable statistical sensitivity.
The direct effect is in our case two orders of magnitudes smaller than the indirect. It is implicitly accounted for in the analysis described in Sec.~\ref{sec:CorrStrat}. 
In contrast, the effect from the mercury comagnetometer was and will be a source of a large systematic effect and is calculated as
\begin{equation}
\label{HgFalseEDM1}
d^{\rm false}_{n \leftarrow {\rm Hg}} = \abs{\frac{\gamma_n}{\gamma_{\rm Hg}}} d^{\rm false}_{\rm Hg} =  \abs{\frac{\gamma_n}{\gamma_{\rm Hg}}} \left(-\frac{\hbar \gamma_{\rm Hg}^2}{2 c^2} \langle \rho B_\rho \rangle\right),
\end{equation}
where the angle brackets correspond to the volume average over the precession chamber. 
Injecting the polynomial expansion of Eq.~\ref{eq:HarmonicPolynomialExpansion} into this expression, it becomes
\begin{equation}
d^{\rm false}_{n \leftarrow {\rm Hg}} =
- \frac{\hbar \abs{\gamma_n \gamma_{\rm Hg}}}{2 c^2}
\sum_{l,m} G_{l,m} \langle \rho \Pi_{\rho, l, m} \rangle.
\end{equation}
In case of a cylindrical precession chamber of radius $R$ and height $H$, with the center of the cylinder being the coordinate system origin, only the modes $\Pi_{\rho, l, 0}$ with $l$ odd contribute to the false EDM, which can then be written up to order~$l=7$ as:
\begin{equation}
\label{HgFalseEDM4}
\begin{aligned}
&d^{\rm false}_{n \leftarrow {\rm Hg}} 
 =  \frac{\hbar \abs{\gamma_n \gamma_{\rm Hg}}}{8 c^2} R^2 \left[ G_{1,0} - G_{3,0} \left(\frac{R^2}{2}-\frac{H^2}{4} \right) \right. \\
& +  G_{5,0} \left(\frac{5R^4}{16} - \frac{5R^2 H^2}{12} + \frac{H^4}{16} \right) \\
& - \left. G_{7,0} \left( \frac{7R^6}{32} - \frac{35R^4 H^2}{64} + \frac{7R^2H^4}{32} - \frac{H^6}{64} \right) \right].
\end{aligned}
\end{equation}

\subsection{Transverse inhomogeneity}
\label{sec:trans}
Another effect which is related to magnetic gradients is the transverse inhomogeneity. 
It induces a frequency shift unrelated to the electric field which moves the frequency ratio $\R = f_\mathrm{n} / f_{\rm Hg}$ by a fraction $\delta_T$ from its unperturbed value $|\gamma_n / \gamma_{\rm Hg}|$.
This effect arises from the difference in the behavior of neutrons and mercury atoms. 
Ultracold neutrons fall into the adiabatic regime of slow particles, $\bar{v}_n\approx \SI{3}{m/s}$, where the typical rate of change of the magnetic field as the neutron crosses the precession chamber is much lower than the Larmor frequency.
Mercury atoms fall into the non-adiabatic regime of fast particles, $\bar{v}_{\rm Hg}\approx \SI{180}{m/s}$, which cross the chamber many times during each precession.
This difference changes the way the particles' spins average the magnetic field, and therefore their precession frequency. While the neutrons' spins effectively average $\left<  \left| \vec{B}\right| \right>$, the mercury spins follow $\left|\left< \vec{B} \right> \right|$. The latter always less than or equal to the former, increasing $\R$.
The expression of the transverse shift is 
\begin{equation}
\label{eqn:quadShift}
\delta_T = \frac{\bt}{2B_0^2},
\end{equation}
where $\bt = \av{(B_x-\av{B_x})^2+(B_y-\av{B_y})^2}$ is the transverse inhomogeneity, which results from field gradients. 
An expression for this in terms of the expansion coefficients $G_{l,m}$ is given in Appendix~\ref{App:Bt2Expression}.

\subsection{Gravitational shift and correction strategy}
\label{sec:CorrStrat}
On top of the transverse inhomogeneity, there are several other effects that can shift the ratio $\R$. 
For the purpose of the present discussion, we write the combination of these effects as 
\begin{equation}
\label{Rshift}
\mathcal{R} = \frac{f_{n}}{f_{\rm Hg}} = \abs{\frac{\gamma_{n}}{\gamma_{\rm Hg}}}
\left( 1 + \delta_{\rm grav} +\delta_{\rm earth}+  \delta_{\rm T} + \delta_{\rm other} \right). 
\end{equation}
The terms correlated to the electric field are not taken into account in this expression.
We have already discussed the $\delta_{\rm T}$ shift in Section~\ref{sec:trans}, while the shift $\delta_{\rm earth}$ arises from the fact that the experiment was performed in the  rotating frame of the earth and is not related to the inhomogeneity of the magnetic field, see Appendix~\ref{sec:rotation}.
The last term, $\delta_{\rm other}$, accounts for small ($<10^{-29}e$~cm) shifts unrelated to field uniformity that are discussed in Table I of~\cite{Abel2020_2} and will not be detailed here.
The first term, $\delta_{\rm grav}$, is the dominant shift in Eq.~\ref{Rshift} and is called the gravitational shift. 
It is caused by the different centers of mass of ultracold neutrons and mercury atoms,
\begin{equation}
\label{eq8}
\delta_{\rm grav} = \pm \frac{G_{\rm grav} \langle z \rangle}{|B_0|}.
\end{equation}
The sign $\pm$ refers to the direction of the magnetic field $B_0$, $G_{\rm grav}$ is the so called gravitational gradient and $\langle z \rangle$ is the relative shift in the center of mass of the neutrons with respect to the mercury, which is significantly non-zero and negative: $\langle z \rangle = -\SI{0.39(3)}{cm}$ \cite{Abel2020_2}. Note, that the center of mass of mercury vapor coincides with the center of the precession chamber; its gravitational offset is negligible.
The term $G_{\rm grav}$ depends on the difference of the magnetic field averaged by both populations and is a function of the gradients $G_{l,0}$ with $l$ odd. 
Details about the calculation of that term can be found in~\cite{Abel2019}.
It is based on the approximation of a neutron density linear in $z$ in the precession chamber.
With a field expansion up to order 7, the expression of $G_{\rm grav}$ is given by the following combination:
\begin{equation}
\label{Ggrav}
\begin{aligned}
&G_{\rm grav} = \left[
G_{1,0} + G_{3,0} \left( \frac{3H^2}{20} - \frac{3R^2}{4} \right) \right.\\
&+ G_{5,0} \left(\frac{3H^4}{112}-\frac{3R^2H^2}{8}+\frac{5R^4}{8} \right) \\
&+ \left. G_{7,0} \left( \frac{H^6}{192} - \frac{9R^2H^4}{64} + \frac{21R^4H^2}{32} - \frac{35R^6}{64}\right)
\right].
\end{aligned}
\end{equation}
The strategy to correct the motional false EDM using the gravitational shift is explained in~\cite{Abel2019} and its application is detailed in~\cite{Abel2020_2}. 
It is an extension of the method used in~\cite{Pendlebury2015} and it will be briefly summarized hereafter. 
We fixed a magnetic-field configuration with a chosen gravitational gradient $G_{\rm grav}$ applied, varied for each sequence of measurements. 
A sequence was a series of consecutive measurements of the neutrons' precession frequency with a nominally fixed magnetic-field configuration. An (anti-)parallel electric field was applied in an ``ABBA'' pattern consisting of 28 single measurements at one electric field polarity, 8 measurements without electric field, 56 measurements at the opposite polarity, again 8 cycles at $E=0$, and a return to the initial polarity for 28 measurements, with each repetition taking around 10 hours. 
This was done to compensate for any (unintentional) linear drifts in any experimental parameter.
Per sequence, we extracted one value of the measured electric dipole moment and its statistical error. 
The cycles at $E=0$ do not contribute directly to the EDM sensitivity, but are necessary to set operation parameters in a way not biased by the blinding or any $E$-dependant systematic effect. 
This measured EDM is then the sum of the true neutron EDM and the mercury induced false one from Eq. \ref{HgFalseEDM4}:
\begin{equation}
\label{falseEDM_Ghat}
d_n^{\rm meas} = d_n^{\rm true} + \frac{\hbar \abs{\gamma_{n} \gamma_{\rm Hg}}}{8 c^2} R^2 \left( G_{\rm grav} + \G \right),
\end{equation}
where $G_\mathrm{grav}$ is separated out and the residual gradient $\G$ is called the phantom gradient.
It is defined as the sum of odd-$l$ order contributions once the $G_{\rm grav}$ contribution is subtracted:
\begin{equation}
\G = \widehat{G_{3}} + \widehat{G_{5}} + \widehat{G_{7}} + \cdots,
\end{equation}
with
\begin{eqnarray}
\widehat{G_{3}} & = & G_{3,0} \left( \frac{H^2}{10}+\frac{R^2}{4} \right), \\
\widehat{G_{5}} & = & G_{5,0} \left(\frac{H^4}{28} - \frac{R^2 H^2}{24} - \frac{5R^4}{16} \right), \\
\widehat{G_{7}} & = & G_{7,0} \left(\frac{H^6}{96}\!-\!\frac{5R^2 H^4}{64}\!-\!\frac{7R^4H^2}{64}\!+\!\frac{21R^6}{64} \right)\!,
\end{eqnarray}
obtained by subtracting Equation \ref{Ggrav} from Equation \ref{HgFalseEDM4}.
For each sequence, inserting Equation \ref{eq8} in Equation \ref{Rshift}, we also extract the frequency ratio
\begin{equation}
\R = \abs{\frac{\gamma_{n}}{\gamma_{\rm Hg}}} \left( 1 + \frac{G_{\rm grav} \langle z \rangle}{B_0}  +\delta_{\rm earth} + \delta_{\rm T} + \delta_{\rm other} \right).
\end{equation}
We define the corrected quantities $d_n^{\rm corr}$, $\R^{\rm corr}$ to be
\begin{equation}
d_n^{\rm corr} = d_n^{\rm meas} - \frac{\hbar \abs{\gamma_{n} \gamma_{\rm Hg}}}{8 c^2} R^2 \G 
\end{equation}
and
\begin{equation}
\R^{\rm corr} = \R - \abs{\frac{\gamma_{n}}{\gamma_{\rm Hg}}} \left(\delta_{\rm T} + \delta_{\rm earth} \right).
\end{equation}
Using the dependency of $\R$ on $G_{\rm grav}$, one can express a linear dependency between $d_n^{\rm corr}$ and $\R^{\rm corr}$ as follows:
\begin{equation}
\label{eqn:crossingLines}
d_n^{\rm corr} = d_n^{\rm true} + B_0 \frac{\hbar \gamma_{\rm Hg}^2}{8 c^2 \langle z \rangle} R^2 \left( \mathcal{R}^{\rm corr} - \abs{\frac{\gamma_{n}}{\gamma_{\rm Hg}}} \right), 
\end{equation}
where $R$ denotes the trap radius, and $\mathcal{R}^{\rm corr}$ the corrected frequency ratio.
With two sets of points $(d_n^{\rm corr}, \R^{\rm corr})$ for both $B_0$ directions, one can fit both sets with a common and opposite slope. 
At the crossing point $(\R_{\times}, d_{\times})$, we get $d_{\times} = d_n^{\rm true}$ and $\R_{\times} = \abs{\gamma_{n}/\gamma_{\rm Hg}}$, free of the systematic effects described in this section.
Therefore, to obtain the systematic-free value of the EDM, the quantities $\G$ and $\left< B_T^2 \right>$ are required for every EDM measurement sequence. 
These quantities were extracted from magnetic field maps taken during the annual proton accelerator and UCN source shutdown.
$\delta_\mathrm{earth}$ is the same in magnitude for each measurement sequence, with the sign inverting depending on the direction of $B$.
It should be noted that, due to the principle of the crossing point method, the corrections of $\mathcal{R}$ have an impact on the nEDM measurement only if they are different for the two directions of the $B_0$ field.

%

\FloatBarrier

\section{The coil system\label{Sec:NEDMMagneticSetup}}
\subsection{Setup description\label{Sec:MagneticSetup}}
\begin{figure}
\includegraphics[width = \columnwidth]{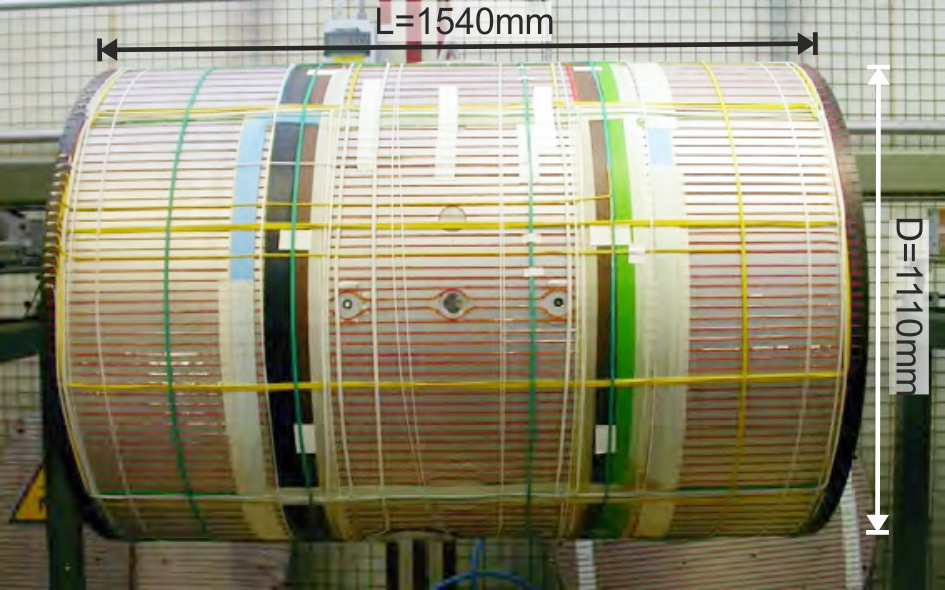}
\caption{\label{Fig:B0Coil} Side view of the $B_0$ coil (red cables) and trimcoils (green, yellow and white cables) wound on the surface of the vacuum tank.}
\end{figure}

As mentioned in Sec.~\ref{sec:Intro}, in order to measure the neutron EDM, a highly uniform magnetic field is required. 
In the PSI experiment, many components were dedicated to the production of such a field and to the reduction of its non-uniformity.
The main coil used to produce the $B_0$ field (called the $B_0$ coil) was a $\cos \theta$ coil of 54 turns wound around the surface of the cylindrical vacuum tank of diameter $D=\SI{1100}{mm}$ and length $L=\SI{1540}{mm}$ (see \autoref{Fig:B0Coil}) to produce a vertical field. 
This coil produced a field with a relative uniformity $\delta B_0/B_0 \sim 10^{-3}$ in the precession chamber, a cylinder of radius \SI{23.5}{cm} and height \SI{12}{cm} with its axis pointing vertically, i.e.,\ along $z$ in \autoref{Fig:simuAnsys}, mounted $+\SI{2}{cm}$ vertically offset from the centre of the coil.
The $B_0$ coil was mounted within a passive magnetic shield. The four layer shield made of mu-metal, a metal alloy with high magnetic permeability, had a quasistatic shielding factor of 1500 to 14000 for small perturbations (smaller than 1~$\muT$), depending on the direction ($x$, $y$ or $z$). 
This factor increases with the amplitude of the perturbation.
Due to the interaction of the field produced by the $B_0$ coil with the innermost layer of the magnetic shield, 40\% of the $B_0$ field came from the magnetization of the shield itself.
As a result, shield imperfections were a potential source of field non-uniformity.
The $B_0$ coil in conjunction with the passive shield generated a $\pm$1~$\muT$ field using a $\pm$17~mA current.

Thirty-three correction coils were used to optimize the magnetic-field homogeneity. They were also wound on the vacuum tank, on top of the $B_0$ coil (see \autoref{Fig:B0Coil}).
A homogenization algorithm, detailed in~\cite{Abel2020}, allowed the calculation of the proper currents for each trimcoil for a given magnetic field configuration (nEDM sequence). 
Several ``guiding'' coils were used to maintain the polarization of the neutrons' spins as they were transported to and from the precession chamber: the non-uniformity they potentially caused had to be taken also into account. 

To keep the ambient external field as stable as possible, we used three pairs of large rectangular coils in a Helmholtz configuration surrounding the experiment.
This system, called the surrounding field compensation system (SFC), added an additional ``active'' shielding factor of \numrange{5}{50} at a bandwidth from \SIrange[]{1}{500}{mHz}.
A feedback algorithm dynamically adjusted the current through each of the six coils using the readings of ten three-axis fluxgate magnetometers positioned near the external layer of the passive shield.
The setup and performance of this system are described in detail in~\cite{Afach2014_sfc}. 
\FloatBarrier

\subsection{\texorpdfstring{$B_0$}{B0} coil simulations\label{Sec:B0CoilSimu}}
\begin{figure}
\includegraphics[width = \columnwidth]{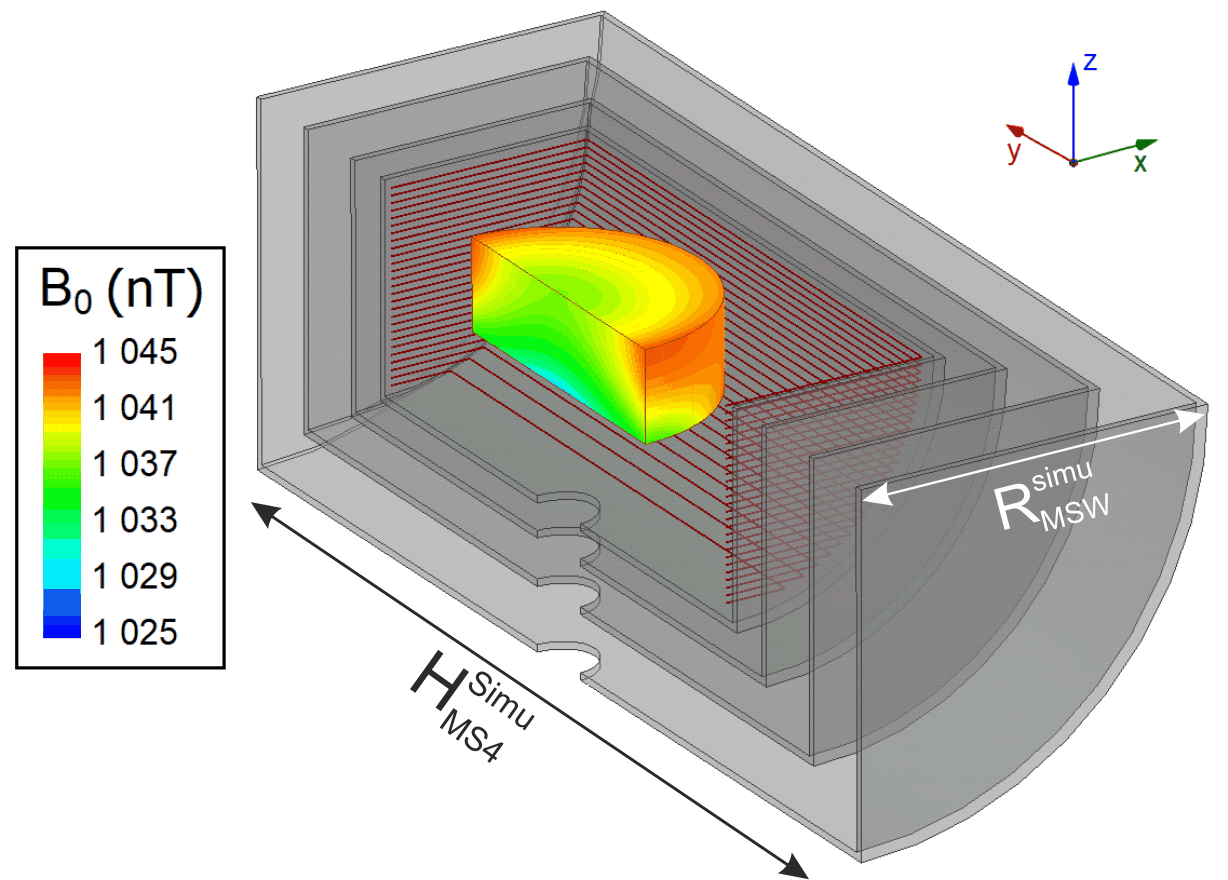}
\caption{\label{Fig:simuAnsys}
Simulation of the field generated by the $B_0$ coil and a 4-layer mu-metal shield. 
The represented geometry is a quarter of the complete volume. 
The external dimensions of the fourth (outermost) layer of the shield were $R_{\rm MS4}^{\rm simu} = 0.98$~m and $H_{\rm MS4}^{\rm simu} = 2.79$~m. 
The coil's windings are represented in red. 
The central volume, the area of the heat map, is a cylinder of diameter 80~cm and height 50~cm, larger than the mapping volume.}
\end{figure}

To validate our understanding of and assumptions about the system, simulations of a simplified geometry of the $B_0$ coil and the passive shield were performed using the Ansys Maxwell software, based on the finite element method. 
A quarter of the simulated geometry and the simulated field is shown in \autoref{Fig:simuAnsys}.
The coil was simulated as a set of 54~independent and perfectly parallel copper loops, with \SI{2}{cm} vertical spacing and a \SI{17}{mA} current flowing through them.
To minimize the computation time, the section of each winding was approximated as a closed rectangle. 
The relative magnetic permeability of the mu-metal composing the shield was set between $\mu =$~10\,000 and $\mu =$~30\,000. 
However, due to the small thickness (\SI{2}{mm}) of the shield layers compared to the scale of the whole simulation, the software had difficulties to generate an adequate meshing and a thicker version of the shield associated with a proportionally smaller value of its permeability had to be used. 
The shield layers were simulated with identical central holes of \SI{20}{cm} diameter along the $z$-axis.

Simulation results and mapping data were analyzed using the same method in order to extract the field gradients (see Sec.~\ref{Sec:SingleMapAnalysis}).
Due to the symmetries of the coil, and an astucious choice of the coordinate system only a few modes of the magnetic field appear.
The first one is the constant term, $G_{0,0}$, which was \SI{1034.47}{nT} in the simulation at the nominal current.
Then, only modes with $l$ and $m$ strictly positive and even appear.
The simulated values and uncertainties for these modes, up to order~6, are listed in Table~\ref{tab:allowedMode}.
The uncertainties were estimated by running several simulations with different parameter settings (meshing refinement, relative magnetic permeability, and proportional changes of the shield thickness).
\begin{table}
\caption{
Ansys simulation values for the magnetic-field modes for a $B_0$ up configuration and their uncertainties.
\label{tab:allowedMode}}
\setlength\tabcolsep{10pt} 
\begin{tabular}{c|cc}
& $G_{l,m}^{\rm simu}$ (pT/cm$^{l}$)
& $\Delta G_{l,m}^{\rm simu}$ (pT/cm$^{l}$) \\
\hline \hline
$G_{0,0}$ & $1034.47 \times 10^{3}$ & $5.08 \times 10^{3}$ \\
$G_{2,0}$ & $-9.26$ & $0.14$ \\
$G_{2,2}$ & $1.18$ & $0.21$ \\
$G_{4,0}$ & $-3.63\times 10^{-3}$ & $0.06\times 10^{-3}$ \\
$G_{4,2}$ & $1.37\times 10^{-3}$ & $0.01\times 10^{-3}$ \\
$G_{4,4}$ & $-8.66\times 10^{-5}$ & $0.14\times 10^{-5}$ \\
$G_{6,0}$ & $-1.16\times 10^{-6}$ & $0.02\times 10^{-6}$ \\
$G_{6,2}$ & $2.77\times 10^{-7}$ & $0.02\times 10^{-7}$ \\
$G_{6,4}$ & $-7.89\times 10^{-8}$ & $0.04\times 10^{-8}$ \\
$G_{6,6}$ & $8.89\times 10^{-9}$ & $0.16\times 10^{-9}$ \\
\hline
\end{tabular}
\end{table}
Although all uneven modes are in principle forbidden, they actually do exist because of the non-perfect geometry of the coil (for example coil connections, cable detours due to holes in the vacuum tank, and non-symmetrical holes in the different layers of the shield). 
It turns out, nevertheless, that they had small amplitudes compared to the even modes.
A comparison between the measured $B_0$ field and the predicted values for these modes will be discussed later.
\FloatBarrier

\section{The magnetic-field mapping}
\subsection{Magnetic field mapper\label{Sec:Mapper}}
\begin{figure}
\includegraphics[width = \columnwidth]{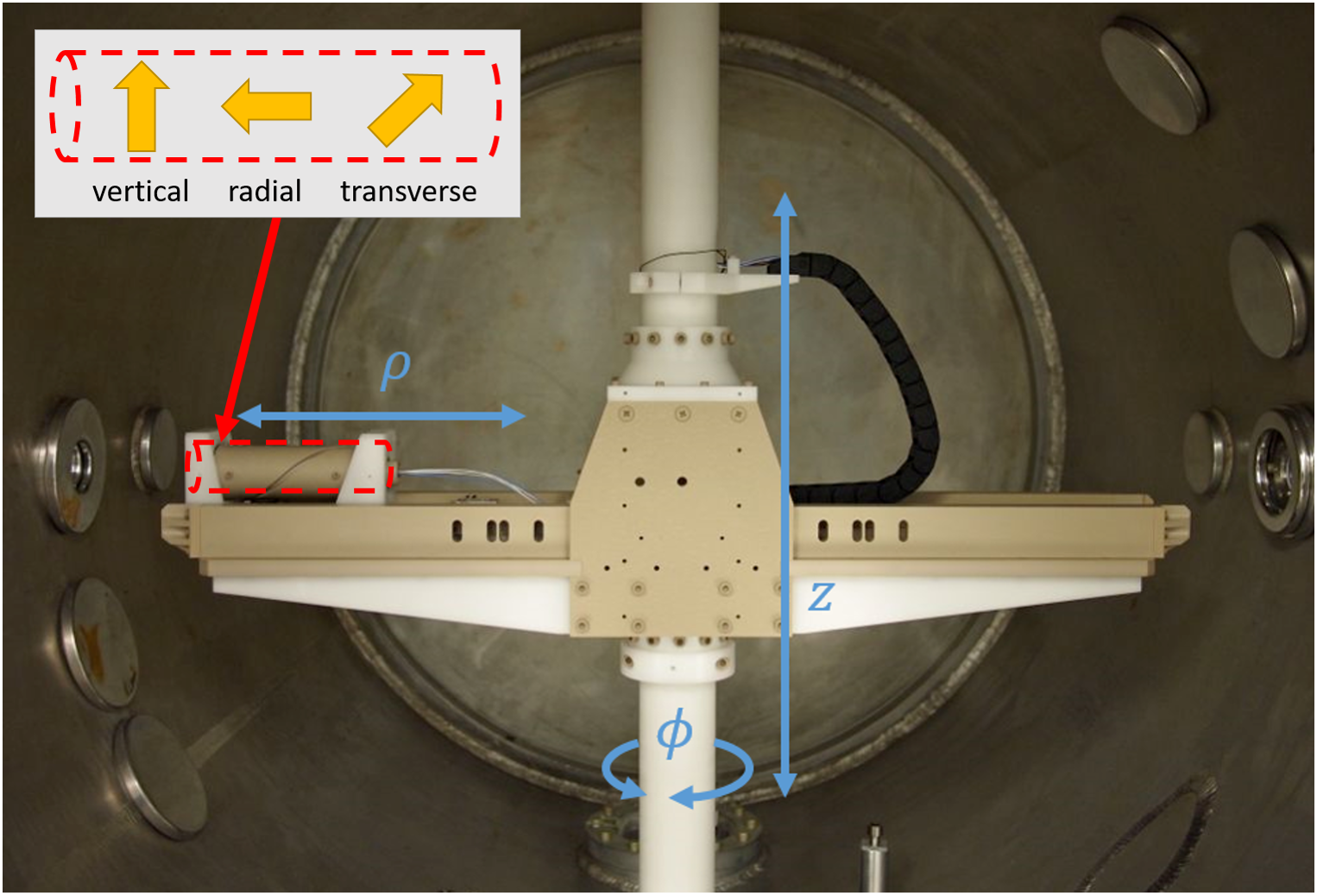}
\caption{\label{Fig:Mapper}
Magnetic-field mapper installed in the empty vacuum vessel. The fluxgate is inside the tube on the left, on which the helical groove used for the calibration motion can be seen. The insert illustrates the relative position of the three individual fluxgate sensor axes, which are offset from each other by \SI{20}{mm} in the radial direction.
}
\end{figure}

The offline magnetic-field characterization was performed regularly during the annual accelerator shutdown period (see~\ref{Sec:MappingCampaigns}) using an automated magnetic field measurement device, the so-called mapper.
This mapper was installed inside the empty vacuum vessel, i.e., with the electrode stack removed.
It allowed the movement of a precise magnetic sensor inside the vacuum vessel, as shown in \autoref{Fig:Mapper}.
The fully-sampled measurement volume was a cylinder of diameter 68~cm and height 32~cm.

The three stepper motors used for the sensor motion along the $\rho$, $\phi$ and $z$ axes were located below the vacuum vessel, outside the cylindrical magnetic shield.
Every part of the mapper inside the magnetic shield was made of non-magnetic materials (PEEK, POM, aluminum, ceramics, glass, etc.\ ), with all materials screened for magnetic contamination in dedicated measurements using a sensitive SQUID (superconducting quantum interference device) magnetometer array at the Berlin magnetically shielded room 2 (BMSR-2) at the Physikalisch Technischen Bundesanstalt~(PTB), Berlin. 
No conductive surfaces were located close to the fluxgate sensor.
This precaution avoided both eddy currents induced by the fluxgate excitation pulses and Johnson noise. 

The $z$ motion was performed using a linear column coupled with a linear transducer, shifting up or down the whole assembly from below.
The $\phi$ motion was done by rotating the central axis of the mapper about a pair of bearings mounted on flanges at the top and bottom of the vacuum tank. 
Finally the $\rho$ motion was performed using a rack and pinion connected to a vertical axle within the lower shaft (coupled to the $\rho$~motor) and to the cart holding the sensor.
The cart was guided along the main plate using twelve non-metallic radial bearing assemblies rolling against linear tracks to constrain all undesired motion.

The $z$-axis position was read with a linear transducer. Although the $\phi$ and $\rho$ positions could be read using wire potentiometers, the best accuracy was provided by counting motor steps in an open-loop fashion.
%
The sensor cart could hold two different sensors:
\begin{itemize}
  \item a low-noise three-axis fluxgate magnetometer,
  \item a two-axis inclinometer (KELAG KAS901-51A).
\end{itemize}
As the inclinometer was slightly magnetic, it was only used to perform mechanical characterisation of the mapper and was removed during magnetic map measurements.

The fluxgate used was a FL3-2 from Sensys, see \autoref{tab:fluxgateProperties}, with three independent single axis detectors mounted along the $\rho$-axis spaced by 20~mm as shown in the insert of \autoref{Fig:Mapper}. 
The specifications for our fluxgate are listed in Table~\ref{tab:fluxgateProperties}. 
\begin{table}
    \caption{Manufacturer specification of the mapper fluxgate (Sensys FL3-2)    }
    \label{tab:fluxgateProperties}
    \begin{tabular}{c|c}
        Characteristic & Value \\
        \hline \hline
        Measurement range & $\pm2~\muT$ \\
        Accuracy & $\pm 0.5$~\% \\
        Orthogonality & $<0.5^\circ$ \\
        Zero drift & $<0.1~$nT/K \\
        Scaling temp. coeff. & +20~ppm/K, typ.\\
        Noise & $<20$~pT$/\sqrt{\mathrm{Hz}}$ \\
        Analog outputs & \SI{5.0}{V/\muT} per sensor\\
        \hline
    \end{tabular}
\end{table}
The stated zero drift only accounts for temperature correlations. 
It turned out that for measurements with an accuracy $<1$~nT, other influences, like $1/f$ noise, dominated the signal stability in time. 
We also found zero-offsets of the order 10~nT for all three independent sensors after several years of use and exposure to a variety of conditions.
Sub-nT accuracy could be reached by an \textit{in situ} zero-offset determination done with the fluxgate mounted onto the mapper, using the same electronics including cables and data acquisition system. 
For such a measurement, a special mechanism to rotate the fluxgate tube about the $\rho$-axis was used.
It combined the helical groove on the fluxgate seen in \autoref{Fig:Mapper} with a pneumatically moved index finger within the upper vertical axis.
The next section explains this key feature of the mapper in more detail. 
\FloatBarrier

\subsection{Fluxgate zero-offset determination\label{Sec:FGCalib}}
A frequently used method to find the zero-offset for a magnetic field detector sensitive in only one spatial direction is the measurement of the magnetic field at one point twice, with the measuring direction reversed for the second measurement. 
The magnetic field value is cancelled when the time between the two measurements is short enough that a possible magnetic field change is negligible.
The average value of both field readings is then the zero-offset. 
The accuracy of such a method depends on the accuracy of the rotation angle $\gamma$, which must be exactly \SI{180}{\degree} to reverse the measuring direction. 
The influence of an uncertainty $\Delta \gamma$ is proportional to the background field strength transverse to the measuring direction of the detector.
Therefore, the highest accuracy for the zero-offset is reached when the background field is small and in the direction of the sensitive axis of the sensor. 
In our case, since the mapper did not allow adjustment of the single detectors in the fluxgate in 3D to the maximal and minimal field reading, we used the center of the magnetic shield for the zero-offset determination.
We observed that the absolute value of the field was lowest close to the center when the shield was degaussed without a $B_0$ field applied. 
Indeed, when comparing the zero-offset measured in the absence and presence of a $B_0$ field, we observed a significant effect.
Without correction, the measured apparent zero-offsets of the horizontal field sensors ($\rho$, $\phi$) were of order \SI{1}{nT}, due to the misalignment of the fluxgate axes into the \SI{1}{\micro T} field into the vertical ($z$) direction. Comparison of measurements taken in different field configurations (around \SI{1}{\micro T} at the center of the tank in each direction $x$, $y$ and $z$ successively) allowed the determination of these angles and for this effect to be corrected.

For our fluxgate zero-offset determination, each single detector was moved one by one to the central position and readings were taken for the ``normal" and the ``reversed" fluxgate orientation. 
In the reversed position, the fluxgate was rotated by $\pi$ around the radial axis, inverting the field readings of the transverse and vertical detectors. 
To measure the offset of the radial sensor, and increase the accuracy of the determination for the transverse and radial sensors, measurements were taken every \SI{10}{\degree} for a rotation around the vertical axis, which lead to 36 pairs of opposing field measurements for the transverse and vertical detectors and 18 (measured nominally twice) for the radial one. 
These measurements are combined to give the final determination of each fluxgate offset.
To be able to bring each sensor to the center of the coordinate system to perform this measurement was an explicit design requirement of the system.
The necessary mapper movement time to measure the positions for all three single detectors was about five minutes.
Repeating the zero-offset determination immediately afterwards, in the real use scenario in a vertical \SI{\pm1}{\micro T} field, leads to a reproducibility of about \SI{30}{pT} in the vertical sensor, \SI{50}{pT} in the radial sensor and \SI{350}{pT} in the transverse sensor.
The poorer reproducibility for the transverse axis is due to a small amount of play that developed in the mechanism locking the fluxgate in the normal or reversed orientation during the hundreds of zero-offset measurement cycles taken during the 2017 mapping campaign, resulting in a worsening of $\Delta \gamma$ over time.
This zero-offset measurement is unique to our mapper and could be performed at any time. 
%
The zero-offset determination procedure is in principle immune to any magnetic field of the parts that are rotated together with fluxgate and to the magnetic field of the fluxgate itself. 

\begin{figure}
    \includegraphics[width = \columnwidth]{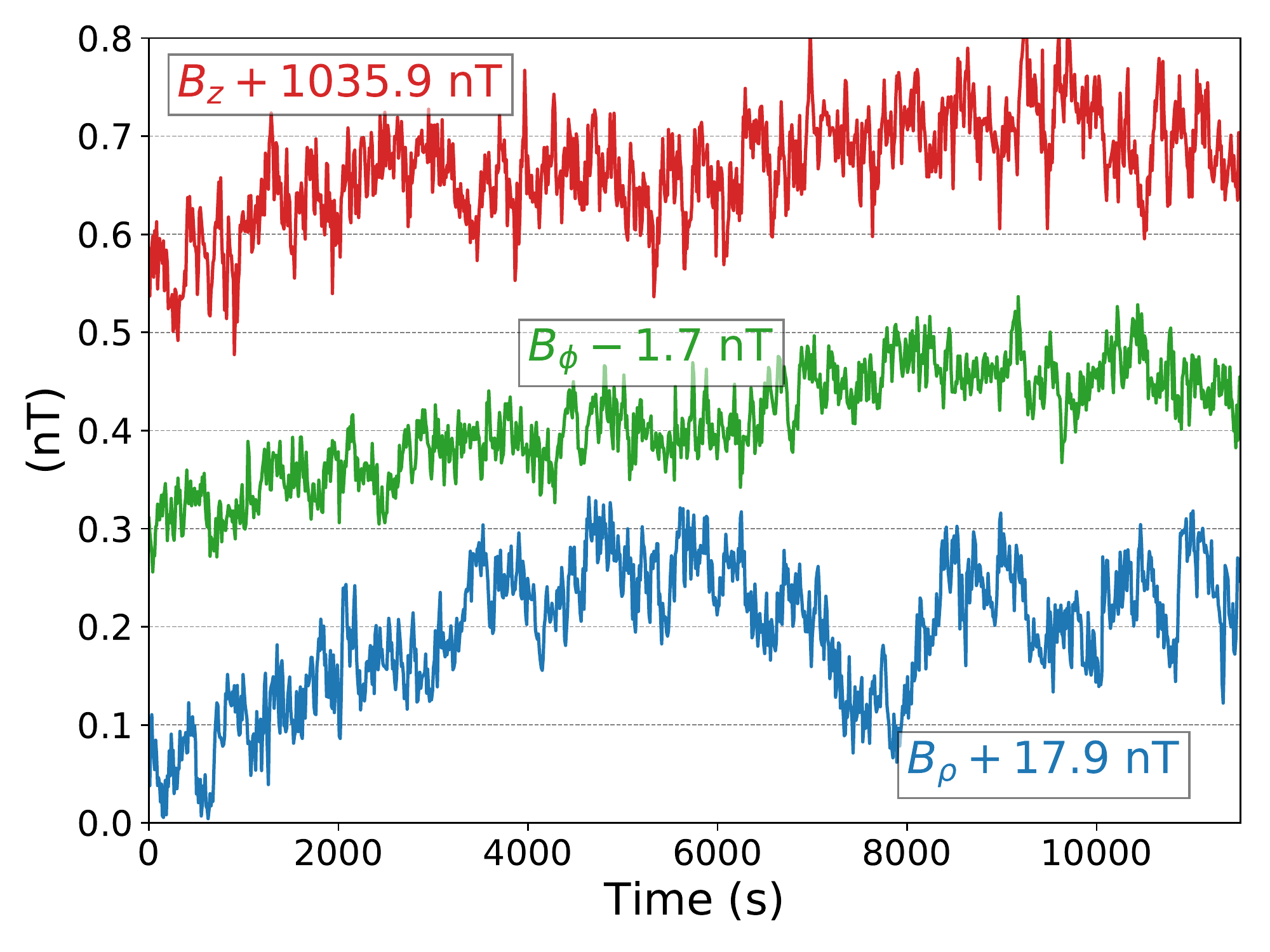}
    \caption{Recording of the field measured by the fluxgate every ten seconds at the center of the coil to see the drifts of the three offsets.}
    \label{Fig:ric}
\end{figure}

\autoref{Fig:ric} shows a typical behavior of the field readings over a time period of \SIrange[range-units=single]{3}{4}{hours} for the three fluxgate channels with the fluxgate motionless at the center of the degaussed shield.
Such measurements were performed regularly during the mapping campaigns.
There is no strong correlation between the different traces and the observed drift is about \SI{300}{pT}.
The temperature around the shield was controlled and stable within $\pm\SI{0.1}{K}$. 
Therefore, the temperature could only account for $\pm\SI{10}{pT}$ (see the zero-drift coefficient in Table~\ref{tab:fluxgateProperties}). 
Magnetic-field drifts as the dominant source could also be excluded by reference measurements with Cs magnetometers. 

\FloatBarrier

\subsection{Mapping campaigns}\label{Sec:MappingCampaigns}
\begin{figure}
    \includegraphics[width = \columnwidth]{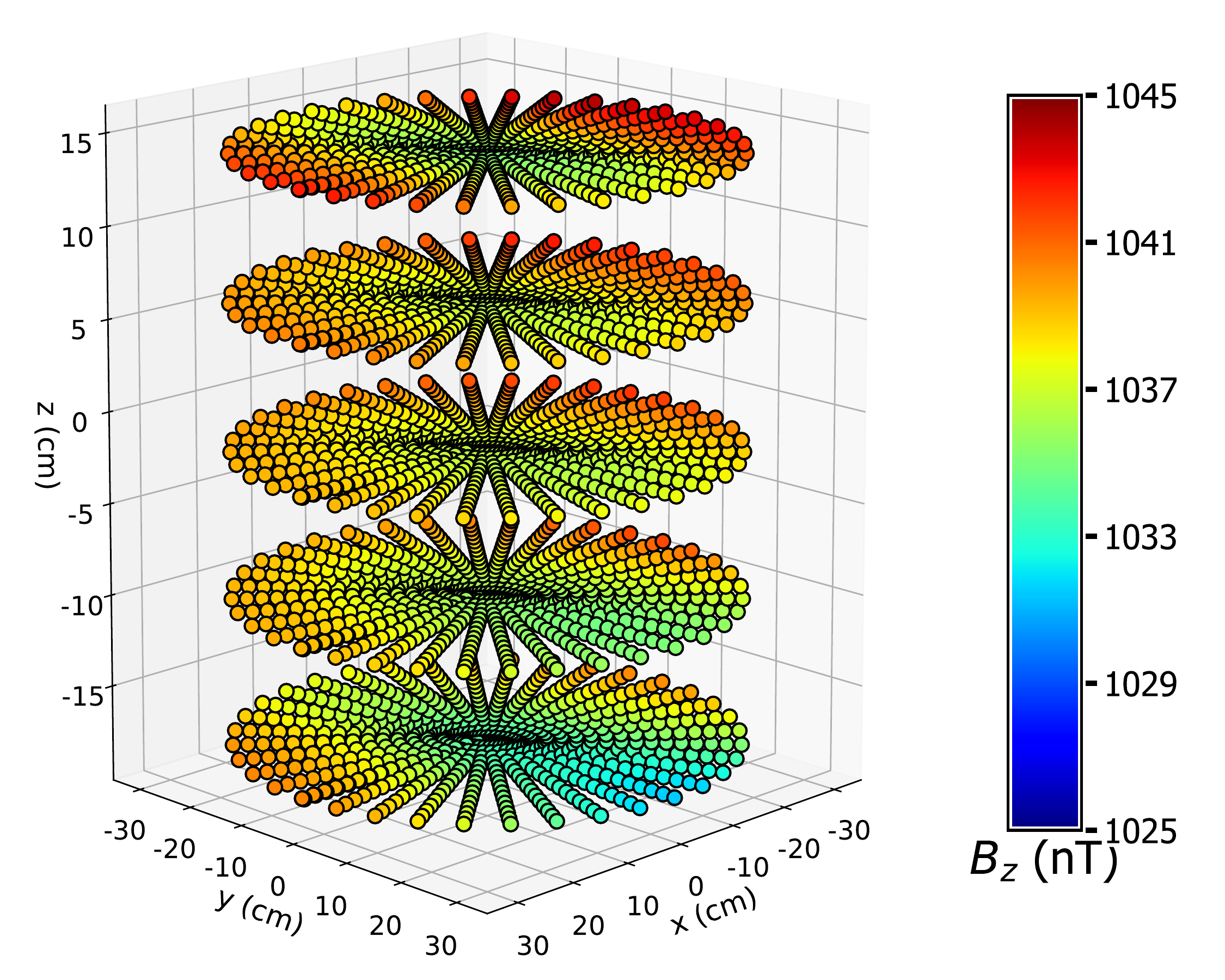}
    \caption{$B_z$ field for a full map of the $B_0$ coil. The axes are defined as in \autoref{Fig:simuAnsys}.}
    \label{Fig:map}
\end{figure}

Three mapping campaigns were conducted in 2013, 2014, and 2017 during which as many as 300 full maps were recorded.
A full map acquisition took between three and six hours.
This time corresponds to a measurement of the vectorial magnetic field for a set of 90 rings (each with 38 points) at 5 given heights (-18, -10, -2, 6 and \SI{14}{cm} in the precession chamber coordinate system, where $z=0$ is at the center of the chamber) and 18 radii (from 0 to \SI{34}{cm}, spaced by \SI{2}{cm} each), as can be seen in \autoref{Fig:map}.
The 2~cm radial spacing between each ring was chosen to match the spacing between the three single-axis sensors contained within the three-axis fluxgate, which is not necessary for the mapping analysis presented here, but useful to obtain a complete 3D representation of the field.
A full map was almost always preceded and followed by one or two zero-offset determination maps to calibrate the fluxgate. 
Moreover, 40-minute recordings of the field were performed at the center of the chamber following each degaussing cycle. 
These recordings allowed us to check the drifts of the fluxgate offsets (see Sec.~\ref{Sec:FGCalib}) and gave time for the fluxgate sensor and the passive magnetic shield to stabilise.

During each mapping campaign, several kinds of maps were taken:
\begin{itemize}
    \item $B_0$ maps, with only the $B_0$ coil powered with $\pm$17~mA.
    \item Maps of the remnant field $B_{\rm rem}$, with all coil currents set to zero.
    \item Trimcoil maps, with only one trimcoil powered with a few mA current.
    \item Guiding coil maps, with only one guiding coil powered with a few mA current.   
    \item Sequence maps, replicating real nEDM measurement conditions. This included powering the trimcoils and guiding coils as they were used during datataking.
\end{itemize}
Each time the state of the $B_0$ coil was changed, the shield was degaussed.
\FloatBarrier

\section{Analysis of a single full map\label{Sec:SingleMapAnalysis}}
In this section, we will describe the analysis method used for a single map.
There were two distinct analysis groups performing differently blinded analyses of the main neutron EDM dataset.
Due to the complex nature of the map analysis and the critical impact it would have on the central value of the reported neutron EDM result, both analysis groups developed independent mapping analyses.
The map measurements supplied to each analysis group were not blinded, but numerical comparisons between the groups were avoided until each analysis was mature and effectively frozen. 
Both analyses were complete and frozen before the unblinding of the main neutron EDM result.
More detailed descriptions of the mapping analyses can be found in \cite{Ayres:2018dbg, Ferraris:thesis}.
In this article we focus on the analysis procedure described in \cite{Ferraris:thesis} and used by the Western analysis group \cite{Ayres_blinding}; the method described in \cite{Ayres:2018dbg} and used by the Eastern analysis group is essentially identical, with the exception that in this analysis the harmonic decomposition described in Subsection~\ref{subsec:HarmonicDecomp} is done using a combined fit for all three axes simultaneously, and a compensation for the radius-dependant misalignment which will be described shortly is explicitly performed (though this was ultimately found to be not necessary when measuring typical nEDM configurations). The results of the two analysis methods ultimately showed excellent agreement.
The positions and the magnetic field will always be given in cylindrical coordinates, as illustrated in \autoref{Fig:Mapper}. The correspondence with the Cartesian coordinate system used in~\cite{Abel2019} and visible on \autoref{Fig:simuAnsys} is the following:
\begin{equation}
    \left\{
    \begin{aligned}
        &\rho = \sqrt{x^2 + y^2}  \\
        &\phi = {\rm arctan}\!\left(y/x\right) \\ 
        &z = z
    \end{aligned}
    \right. .
\end{equation}
An important source of error is the possible misalignment of the \{coil + mapper + sensor axes\} system.
Indeed, if the true vertical axis of the global coordinate system (defined by gravity, and to which the precession chamber is well aligned in normal operation) and the vertical axis of the mapper were not perfectly aligned, or if the angles between the three axes of the fluxgate were not exactly square, the three directions of the field in the chamber would be mixed with each other when measured by the mapper. 
A specific analysis method was developed to reduce the impact of such potential misalignment. 
We measured the vectorial magnetic field. Therefore, we could independently extract the gradients $G_{l,m}$ by analyzing each of the three sensor directions: radial $\hat{r}$, transverse $\hat{\phi}$ and vertical $\hat{z}$.
Let's consider the simple case of a small angle $\alpha$ between the nominal and real axes of the fluxgate, causing a component of the large vertical field to be captured by the radial or transverse sensor.
One can express the vertical and horizontal field mixing effect of such a misalignment as a function of $\alpha$.
Since $\vec{B}_0$ was mainly aligned with the mapper axis along $\hat{z}$ (and the vertical $z$-axis in the global experiment coordinate system), the impact of the horizontal field components $B_{\rm h}$ in the chamber on the measured vertical $z$ field ($B_z^{\rm meas}$) could be neglected. 
The measured vertical and horizontal fields are
\begin{equation}
    \label{Eq:MisalignementImpact}
    \left\{
    \begin{aligned}
        &B_z^{\rm meas} =  B_z \cos\alpha, \\
        &B_{\rm h}^{\rm meas} = B_{\rm h} \cos\alpha + B_z \sin\alpha,
    \end{aligned}
    \right.
\end{equation}
%
where \guil{meas} denotes the field measured by the fluxgate sensor, and \guil{h} stands for horizontal (radial $\hat{r}$ or transverse $\hat{\phi}$)
Since $\alpha$ is small, we can perform a Taylor expansion:
\begin{equation}
    \label{Eq:MisalignementImpact2}
    \left\{
    \begin{aligned}
        &B_z^{\rm meas} =  B_z \left( 1 - \frac{\alpha^2}{2} \right) + \mathcal{O}(\alpha^3), \\
        &B_{\rm h}^{\rm meas} = B_{\rm h} \left( 1 - \frac{\alpha^2}{2} \right) + B_z \alpha + \mathcal{O}(\alpha^3). 
    \end{aligned}
    \right.
\end{equation}
It is obvious that the measured vertical field is much less impacted by a possible misalignment angle $\alpha$. 
This is most relevant when considering field modes with order $m=0$, due to the relatively large size of the $G_{0,0}$ term, corresponding to the target homogeneous vertical field. 
Therefore, to extract the $G_{l,0}$ gradients, only the vertical $z$ sensor's analysis is used.

The analysis of one direction of the field is divided into several steps that we will describe hereafter.
This method was used for all field directions individually.
However, we will detail it in the next sections for the vertical $z$ direction.

\subsection{Ring by ring analysis\label{Sec:RingByRingAnalysis}}
Due to the measurement pattern of a map, a full map can be seen as a set of 90~rings ($\rho$ and $z$ fixed) of 37~points from \SIrange[]{0}{360}{\degree} plus an additional point at \SI{0}{\degree} (see \autoref{Fig:map}).
The first analysis step is analogous to a Fourier decomposition ring by ring.
For one ring $i$, since the radius $\rho_i$ and height $z_i$ are fixed, the magnetic field is simply a function of $\phi$.
We fit it with a Fourier series as follows, using a simple $\chi^2$ fit, with the Fourier coefficients $a_{m,z,i}$ as parameters of the fit,
\begin{equation}
\label{FourierRing}
B_z \left(\rho_i, \phi, z_i\right) = \sum_{m\geq0} \left[ a_{m,z,i}\cos{(m\phi)} + a_{-m,z,i}\sin{(m\phi)} \right],
\end{equation}
where $\rho_i$ and $z_i$ are respectively the radius and the height of the ring $i$.
The 38~points of a ring are treated equally. 
To compute distinct weights for each point, we would need to include the error due to the fluxgate offset drift.
However, we are not able to estimate that error \emph{a priori}.

\begin{figure}
    \subfigure[]{\includegraphics[width =0.95 \columnwidth]{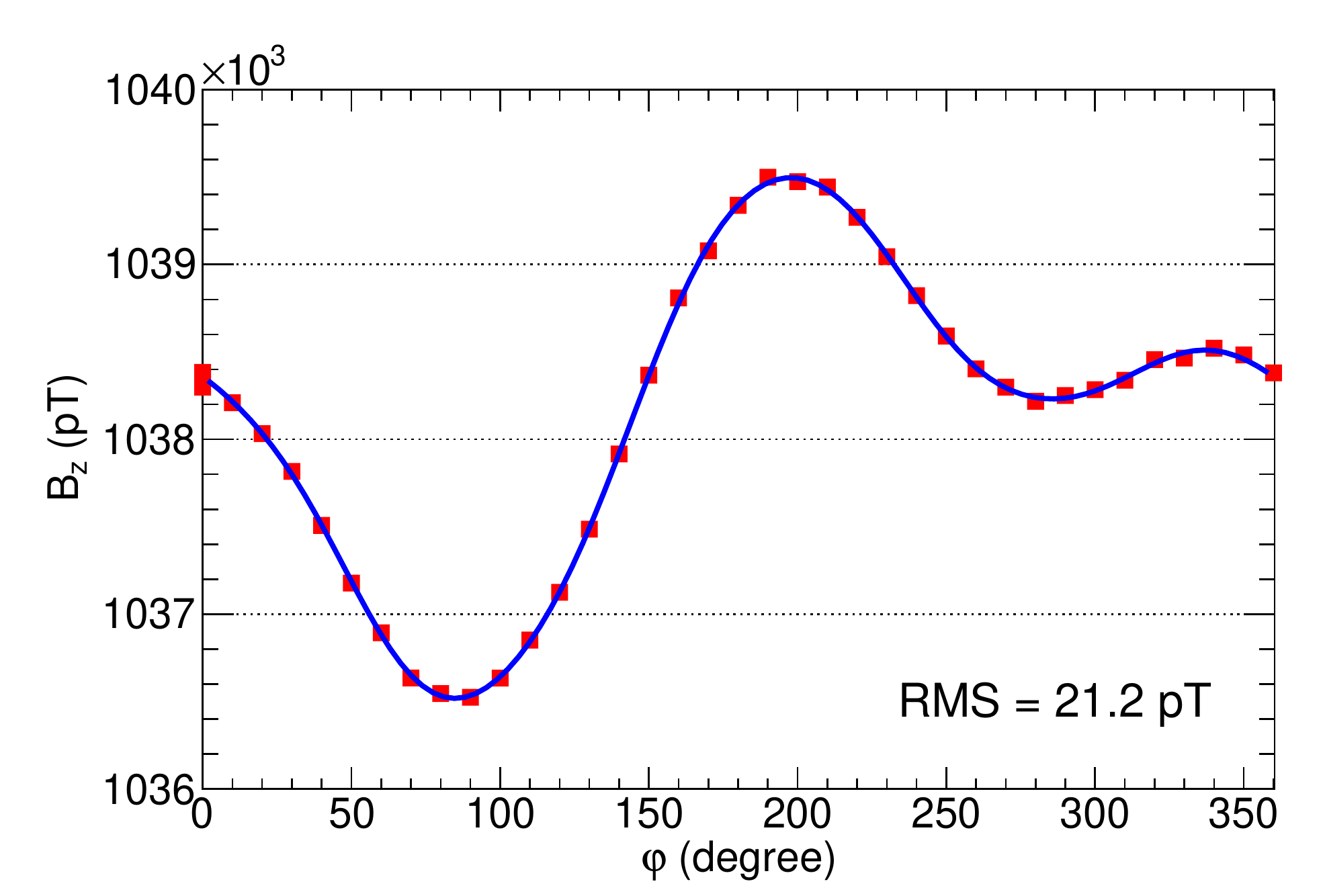}\label{Fig:FourierFitA}}
    \hfill
    \subfigure[]{\includegraphics[width = 0.95\columnwidth]{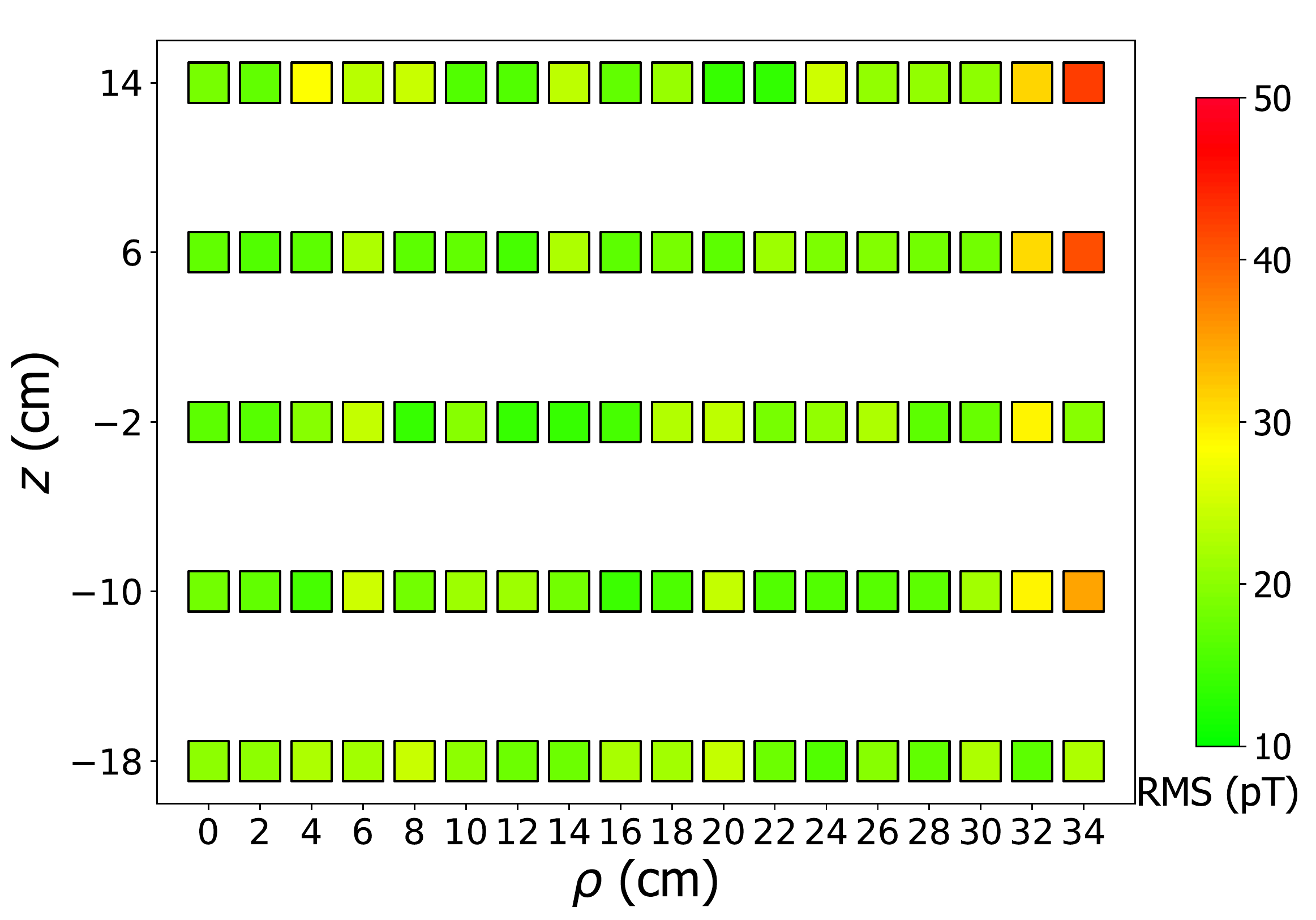}\label{Fig:FourierFitB}}
        \caption{Fit of the $B_z$ field with a Fourier series up to order $m=6$ for a $B_0$ up map. In~(a) is the fit for the ring $\rho=\SI{22}{cm}$, $z=6$~cm. In~(b) are the square root of mean squared residuals of all the rings. 
    Each square corresponds to the RMS residual after fitting the ring at the position $(\rho, z)$. 
    \label{Fig:FourierFit}}
\end{figure}

The Fourier fit step gave us a set of Fourier coefficients $a_{m,z, i}$ per ring $i$ with their associated errors.
These errors were scaled with the factor $\sqrt{\chi^2_i / \mathit{NDF}}$, with $\mathit{NDF}$ the number of degrees of freedom of the fit, to take into account the quality of each ring $i$ for the next analysis step.
An example of this fit for a $B_0$ map can be seen in Figure~\ref{Fig:FourierFitA}. 
The fit was done up to order $|m|=6$ (13 coefficients). 
This limit was chosen for several reasons:
\begin{itemize}
\item The improvement of the fit residuals between order~6 and order~7 was not significant.
\item The contribution of the order $m=7$ to $\G$ was smaller than the reproducibility of the degaussing process. As we will discuss in Section \ref{Sec:linearity}, this is the limiting factor in the correction of nEDM systematics.
\item The contribution of the order $m=7$ to $\bt$ was negligible, being much less than the degaussing reproducibility for this quantity.
\end{itemize}
We can compare the quality of a Fourier fit by looking at the square root of the mean squared residual (RMS residual). 
These residuals are displayed in Figure~\ref{Fig:FourierFitB} for a $B_0$ map.
The average value is around \SI{20}{pT}, which is the same order of magnitude as the variations of the fluxgate output over a time similar to the duration of a ring measurement (80~s). 
One can see that the fits of the external rings tend to be of poorer quality.
That may be explained by the higher order terms which grow very quickly at larger distances to the center and are not so well fitted.  
This effect is taken care of by de-weighting the external rings in the next step of the analysis.
\FloatBarrier

\subsection{Harmonic decomposition of the Fourier coefficients\label{subsec:HarmonicDecomp}}
After having extracted a set of Fourier coefficients for each ring $i$, the second step of the analysis is to fit these coefficients with the harmonic functions of the field expansion.
Since we already took care of the $\phi$-dependency of the field by fitting the rings, we will now fit the coefficients with the expansion functions (see Eq.~\ref{eq:HarmonicPolynomialExpansion}) also freed from this dependency.
As mentioned in Sec.\,\ref{Sec:SystematicEffects}, these functions can be expressed as the product of a polynomial in $(\rho, z)$ and a trigonometric function in $\phi$.
As an example, in the case of the $z$ direction:
\begin{equation}
    \label{eq:RedefPolynomials}
    \Pi_{z,l,m}\left(\vec{r}\right) = \left\{
    \begin{aligned}
    &\widetilde{\Pi}_{z,l,m}\left(\rho,z\right) \times \cos \left(m\phi\right) \ {\rm for } \ m\geq0, \\
    &\widetilde{\Pi}_{z,l,m}\left(\rho,z\right) \times \sin \left(m\phi\right) \ {\rm for } \ m<0.
    \end{aligned}
    \right.
\end{equation}
We exploit this property of the harmonic functions when expressed in cylindrical coordinates to fit the Fourier coefficients.
The coefficient $a_{m,z}$ is fitted with a linear combination of the $\widetilde{\Pi}_{z,l,m}$ for different values of $l$, with the order $m$ being the one related to the $\phi$-dependency.
Similarly, $a_{-m,z}$ is fitted with a linear combination of $\widetilde{\Pi}_{z,l,-m}$.
Due to our choice of basis fields, there is no ``mixing'' between terms of different $m$ (i.e. different $\phi$-dependence).
The fit of every Fourier coefficient of a given order $\pm m$ can then be written as
\begin{equation}
    \label{Eq:HarmonicFit}
    a_{m,z,i} = \sum_{l\geq0} G_{l,m} \ \widetilde{\Pi}_{z,l,m}\left(\rho_i,z_i\right).
\end{equation}
This can be compared to Equations \ref{eq:HarmonicPolynomialExpansion} and \ref{FourierRing}.
For the Fourier fit, we use a $\chi^2$ minimization.
There are as many fits as the number of Fourier coefficients extracted from each ring in the first step of the analysis.
\begin{figure}
    \includegraphics[width = \columnwidth]{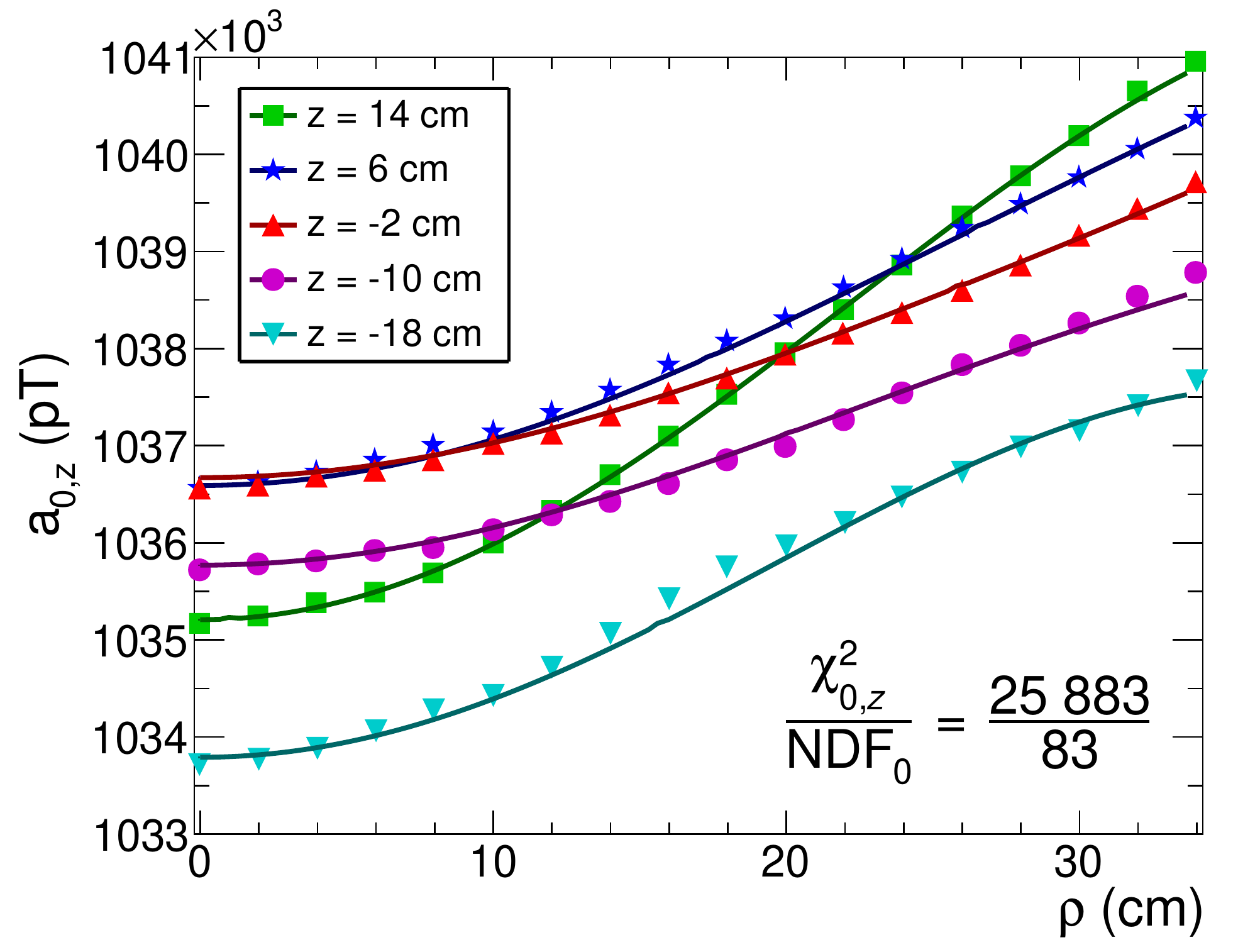}
    \caption{Fit of gradients $G_{l,0}$ to the Fourier coefficients $a_{0,z}$ for a $B_0$ up map. The index $m=0$ denotes the field components without $\phi$-dependence, which are responsible for the ``phantom'' fields, contributing to $\G$. The colors represent the different values of the ring's height $z$, for the same fit. Each point represents the fitted $a_{0,z}$ of a ring. Error bars are too small to be visible.}
    \label{Fig:harmonicFit}
\end{figure}
On \autoref{Fig:harmonicFit}, an example of such a fit is shown. 
This is the fit of the order $m=0$, which gives us the gradients $G_{l,0}$ used to calculate $\G$.
The harmonic fits are performed up to order $l=6$, for the same reasons as used to justify our choice of the largest $m$ in the ring fit stage.

After the harmonic fits, we obtain 60, 54, and 49 gradients and their associated errors, respectively for the analysis directions $\rho$, $\phi$ and $z$.
As can be seen in Tables~\ref{adequateCylindrical} to~\ref{adequateCylindrical3} in the Appendix, this difference in the number of extracted coefficients is due to some gradients not producing a signal in all 3 dimensions.
For fits such as the one in \autoref{Fig:harmonicFit}, the Fourier coefficients' error bars are underestimated, therefore the values of the $\chi^2$ are quite large. 
It turns out that this underestimation of the error bars was due to the drifts of the fluxgate's offsets.
As said in Sec.~\ref{Sec:FGCalib}, the drift of these offsets was approximated as linear, but it can be seen in \autoref{Fig:ric} that this was not always true.
During the small duration of a ring measurement ($\sim\SI{80}{\second}$), the impact of the drifts was very limited and the errors coming from the Fourier fits were therefore not impacted.
However, from one ring to another, with the recording of one map taking several hours, this impact became visible in the terms with $m=0$. 
To take this into account, the Fourier coefficient errors and therefore the gradients errors were scaled with the factor $\sqrt{\chi^2 / \mathit{NDF}}$, with $\mathit{NDF}$ the number of degrees of freedom of this fit.
The phantom gradient $\G$ is calculated directly at this step for the $z$ direction, since it is a linear combination of gradients $G_{l,0}$ which come from the same fit and are therefore correlated.

The last step of the analysis is the combination of the three analysis axes, except for order $m=0$, which is obtained from the analysis of $B_z$ only.
This combination is a simple weighted average of all axes (when available) for each gradient.
After this combination, for one map, we get the 61 gradients $G_{l,m}$ that we use to calculate $\bt$.
The uncertainties obtained from that analysis then take into account the fit error and the non-linear drifts of the fluxgate's offsets. 
In the next section, we present an overview of the systematic errors of the mapping and their impact on the gradients.

\subsection{Systematic errors\label{subsection:syst}}

A variety of additional effects may bias the results of the mapping, arising from mechanical imperfections in the construction and installation of the mapper device.
A few specific classes of errors were considered.

Firstly, the guiding rails along which the mapper cart moved radially were found to be warped. 
This resulted in the misalignment of several milliradians of the radial and transverse sensors into the vertical direction. 
As such, the large vertical magnetic field ($B_z \approx \SI{1}{\muT}$) caused large radius-dependent false fields of several nanotesla in these sensors. 
This observation was validated by separate measurements using an inclinometer mounted at the same position as the fluxgate, as well as direct measurements of the rail profile using a coordinate measurement machine. 

This type of misalignment does not depend on $\phi$, thus the most affected field components are those with $m=0$ due to the predominance of $G_{0,0}$ over all other terms. 
Such false fields do not satisfy the Maxwell equations, therefore the field decomposition basis chosen cannot describe them. 
Thus, in order to evaluate the misalignment in-situ, a fit of the magnetic fields described by $G_{0,0}$ to $G_{6,0}$ and two misalignment angles $\alpha$ and $\beta$ (describing a rotation of the fluxgate about its $\hat{r}$ and $\hat{\phi}$ axes respectively)  for each radius $\rho$ to the Fourier coefficients $a_{0,\rho,\phi,z}$ was performed. 
These results were compatible with the results of the inclinometer measurements and the measurement of the rail profile, and the values of $G_{l,0}$ obtained were compatible with those returned by the main analysis pathway detailed in \autoref{Sec:SingleMapAnalysis}. 
It was found that ignoring the components $a_{0,\rho}$ and $a_{0,\phi}$ in the standard analysis pathway was sufficient to result in unbiased results with comparable precision, while substantially reducing the complexity of the analysis.

Secondly, each of the three fluxgate sensors is specified to be aligned along the nominal direction with a tolerance of \SI{0.5}{\degree}. 
In our case, trying to measure inhomogeneities in, and small transverse components of, a large vertical field, this could have also caused undesirable false fields to appear in the radial and transverse fluxgate sensors, on the order of nanotesla. 
Again the predominant contribution comes from the large $G_{0,0}$ component.
These false fields are then approximately constant for each magnetic-field configuration, meaning they do not cause errors in the estimation of the gradients. 
However, they have to be considered in the analysis of fluxgate zero-offset determination sequences performed in an offset field if absolute values independent of the applied magnetic field are required.
Additionally, these angles become relevant when taking a map of a magnetic field with the largest component in the (horizontal) $\hat{x}$ or $\hat{y}$ direction.

Inaccuracies in the mapper positioning could also lead to measurement error. Although each small stepper motor step corresponded to a high positioning resolution of at least \SI{50}{\micro m}, the real world performance was not proven to this level. Deviation from linearity, a scaling error, or some statistical error in the $\phi$ position would lead to a poor fit at the stage of the ring-by-ring Fourier fit. In the case of the $\rho$ and $z$ positions, a poor fit would be observed at the next step when the gradients $G_{l,m}$ are fitted to the coefficients $a_{m}$. The goodness of fit in real data in the fits for terms $m \neq 0$ was sufficient to exclude such systematic effects at a relevant level, and the measurement uncertainty for the terms $m=0$ much better explained by the aforementioned fluxgate drifts.
Moreover, the zero position of each of the three axes was relatively difficult to determine and accurate to only approximately \SI{0.5}{mm}. %
In the case of the rotational axis of the mapper, any zero-position error of $\phi$ will not affect the values obtained for $\langle  B_T^2 \rangle $ or $\G$ due to the cylindrical symmetry of the precession chamber.
However, a zero-position error on the radial $\rho$ or vertical axes $z$ could cause an anomalous reading, without substantially impacting the goodness of fit.

To estimate the magnitude of this effect, the full analysis of several real maps was repeated while adding a small offset onto each recorded position. Considering the $\rho$-coordinate, it was found that adding an offset of \SI{+1}{mm} to all points for a $B_0$-up map would typically lead to a shift of around $+\SI{0.04}{pT/cm}$ in $\G$ and $-\SI{0.02}{nT^2}$ in $\langle B_T^2 \rangle$. Uncorrected, both lead to a systematic shift in the measured nEDM value of less than $\SI{2e-28}{\elementarycharge~cm}$.

When the correction strategy described in \autoref{sec:CorrStrat} is used to correct the mercury induced false neutron EDM systematic effect described in \autoref{sec:hgIndFalseNEDM}, data taken with both $B_0$ field directions are combined. 
Two lines with the same gradient but opposite sign are fitted to the corrected data of each field direction respectively following Equation \ref{eqn:crossingLines}. This yields the ``crossing lines'' pictured in Figure 4 of \cite{Abel2020_2}. One can imagine that some systematic error like an error in the measurement of $\bt$ shifting $\R$ in the same direction for both signs of B will not affect the crossing point $d_X$ which gives the final corrected nEDM value $d_n^{\mathrm{true}}$, only the crossing point $\R_X$ will be shifted. Meanwhile, some error causing a false EDM reading, for example a systematic error in the determination of $\G$, will only cancel if the sign of the error is opposite for opposite polarities of B.

Considering a $B_0$-down map, we find the same values and same signs for the same $\rho$-offset as for a $B_0$-up map.
Thus, any effect on $\langle B_T^2 \rangle$ is cancelled implicitly when evaluating the neutron EDM. However, the crossing point $\R_{\times}$ would be affected. The value arising from the neutron EDM crossing lines analysis was compatible with a previous, independent determination by the collaboration~\cite{Afach2014}.
There is no such cancellation in this case for the measurement of $\G$. Both signs of B will measure an EDM shifted in the same direction.
The effect on the measured neutron EDM will then be less than $\SI{2e-28}{\elementarycharge~cm}$ per mm of offset in the $ \rho$ value, due to the effect on $\G$.

For shifts in the $z-$position, a similar effect can be observed. For both $B$-field directions, up and down, injecting an offset of $+\SI{1}{\milli\meter}$ leads to a $+\SI{0.2}{pT/cm}$ shift in $\G$, and a shift in $\left< B_T^2 \right> $ of $+\SI{0.02}{nT^2}$.
Again, although the effect on $\left< B_T^2 \right>$ will cancel, the final measured nEDM value would be shifted by \SI{1e-27}{\elementarycharge~cm} for an offset of \SI{1}{mm}.
We conservatively estimate \SI{1}{mm} to be the upper bound on any such shift in the zero-position of the $\rho$ and $z$ axes, leading to an upper bound on the final nEDM systematic error due to this effect of less than \SI{1e-27}{\elementarycharge~cm}.


\section{Global analysis\label{Sec:GlobalAnalysis}}
\begin{figure}
\includegraphics[width = \columnwidth]{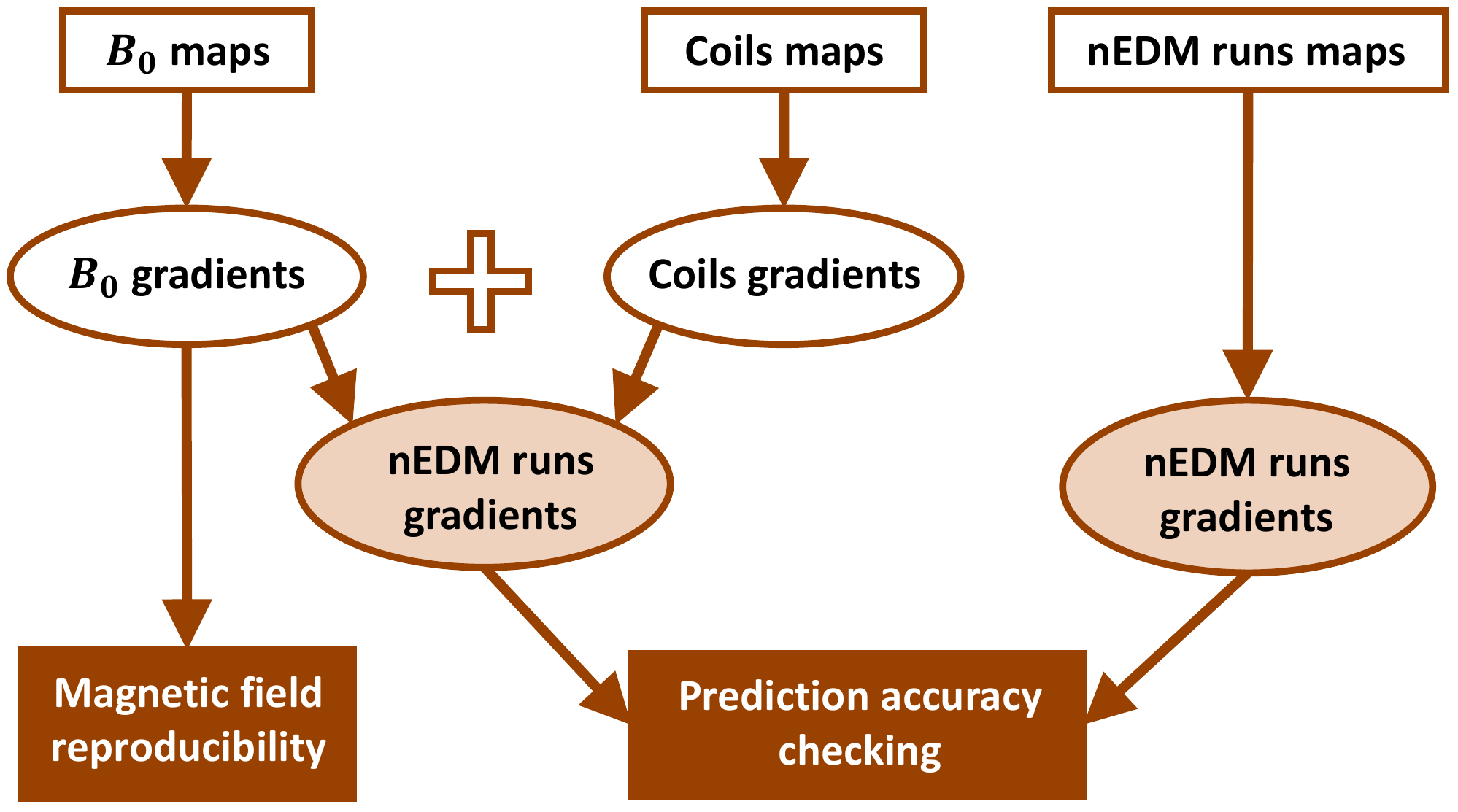}
\caption{\label{Fig:globalAnalysis}
Principle of the global analysis of all maps.
}
\end{figure}

Different kinds of maps were taken during the mapping campaigns.
Each individual map was analyzed with the method described in the previous section to obtain the magnetic-field gradients.
However, to check the quality of the maps and therefore the reliability of the extracted gradients, a global analysis of all maps was performed. 
A schematic diagram of the global analysis is shown in \autoref{Fig:globalAnalysis} and its different parts will be discussed in the following sections.

The field reproducibility and the mapping repeatability were extracted from the global analysis of all the $B_0$ maps.
These two sources of uncertainty of the mapping are defined and discussed in the following section.
With these maps we also extracted the contribution of the $B_0$ coil to the field gradients in the precession chamber. 

Using the trimcoil and guiding coil maps, we had measured the contribution to the gradients of each individual additional coil. 
By combining the $B_0$ coil gradients and those of the other coils, we obtained a prediction of the field gradients for any magnetic configuration.

\subsection{Reproducibility and repeatability\label{Sec:ReprodAndRepeat}}
Two important quantities to evaluate the mapping uncertainties are the field reproducibility and the mapping repeatability. 
The field reproducibility quantifies how reproducible the magnetic state of the system is after applying a standardized degaussing process.
The mapping repeatability, on the other hand, represents our ability to measure twice the same field with the mapping and its analysis without changing any magnetic conditions (no degaussing, identical currents, etc.).
A poor repeatability can be caused by measurement imperfections such as drifts in time of the sensor offset or small variances in misalignment angles. 
In principle, the analysis method aims to take such imperfections into account.
Unlike the reproducibility, the repeatability can therefore be improved by improving either the measurement method or the analysis.

As the $B_0$ field was the main contribution to the magnetic field, only $B_0$ maps were considered to evaluate the reproducibility and the repeatability.
In the following, we will only describe the extraction method for the gradient $\G$, the method being independent of any particular gradient.
To extract the $\G$ reproducibility, during each campaign, several groups of $B_0$ maps were recorded, with a degaussing of the shield in between two groups. Each group itself consisted of a series of $B_0$ maps taken without degaussing in between.
The fluctuations of the measured $\G$ between the different groups quantify the reproducibility. 
However, a na\"ive method would be influenced by the repeatability, which is responsible for the fluctuations of the measured $\G$ between successive maps.
The repeatability was estimated by studying the $\G$ fluctuations within a group.
Both the field reproducibility and the mapping repeatability were found to be independent of the polarity of the field. Therefore, they could be extracted by considering all $B_0$ maps (taking account of the different central value for different polarities).

Due to the different sizes of the groups, one to three maps per group, there was no standard statistical model to estimate the reproducibility $\sigma_{\G}$ and the repeatability $\tau_{\G}$.
Therefore, we derived estimators depending on the number and size of the groups. 
First, we define the estimator of the average gradient of a group $i$ containing $n_i$ maps,
\begin{equation}
\label{Eq:groupAverage}
\overline{\G}_i = \frac{1}{n_i} \sum_{j=1}^{n_i} \G_{ij}.
\end{equation}
Then, using all groups average values with the deviation of the gradient inside each group, we estimate the repeatability as
\begin{equation}
\label{Eq:repeatability}
\tau_{\G}^2 = \frac{1}{N-g} \sum_{i=1}^{g} \sum_{j=1}^{n_i} \left( \G_{ij} - \overline{\G}_i \right)^2,
\end{equation}
where $N$ is the total number of maps and $g$ is the number of groups.
With the group averages, we also estimate the global average value of the gradient produced by the coil $B_0$.
This global average will be useful to predict the gradient of a magnetic configuration and is defined as
\begin{equation}
\label{Eq:globalAverage}
\overline{\overline{\G}} = \frac{1}{N} \sum_{i=1}^{g} n_i \overline{\G}_i 
= \frac{1}{N} \sum_{i=1}^{g} \sum_{j=1}^{n_i} \G_{ij}.
\end{equation}
Finally, from the deviation between the global and individual averages, one can extract the reproducibility and subtract the repeatability contribution as follows:
\begin{equation}
\label{Eq:reproducibility}
\sigma_{\G}^2 = \frac{1}{g} \sum_{i=1}^{g} n_i \left( \overline{\G}_i - \overline{\overline{\G}} \right)^2 - \tau_{\G}^2.
\end{equation}

\begin{figure}
    \includegraphics[width = \columnwidth]{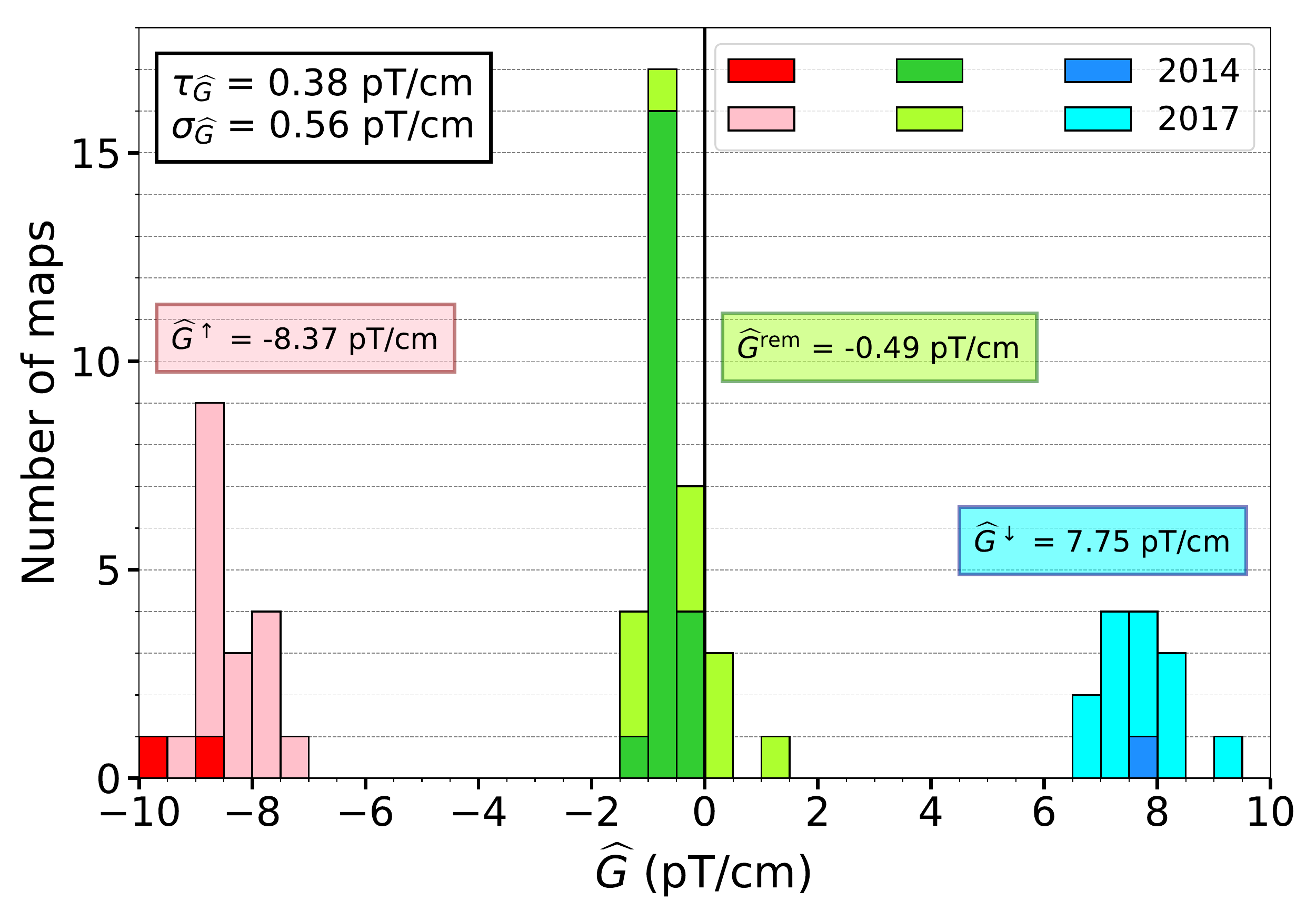}
    \caption{Histogram of the values of $\G$ and its global averages for all the $B_0$ up (red) and down (blue) maps and the remnant field (green) maps. The reproducibility and repeatability were only calculated with the $B_0$ maps.}
    \label{Fig:reproducibility}
\end{figure}

The results of the mapping for the phantom gradient $\G$ and the spread of the measurements can be seen on \autoref{Fig:reproducibility}. 
No maps from the 2013 campaign and only a part of the 2014 campaign maps were used to correct the nEDM data or for this meta-analysis.
This was due to the presence of magnetic elements within the shield which were removed during the 2014 campaign.
The decision to not use the maps recorded before the removal of those elements was taken to avoid any bias in the gradient estimations.
However, it should be said that these maps were studied, too, and confirm an excellent reproducibility of the phantom gradient over the duration of the different campaigns.
The field reproducibility and mapping repeatability were found to be $\sigma_{\G} = \SI{0.56}{pT/cm}$ and $\tau_{\G} = \SI{0.38}{pT/cm}$, respectively. Note that the phantom gradient produced by the $B_0$ coil was very symmetric about zero in up and down configurations as shown in Figure \ref{Fig:reproducibility}. 
For the remnant field, this gradient was close to zero, which was not the case for all the field coefficients.
For the other quantity of interest, the transverse inhomogeneity $\langle B^2_T \rangle$, the reproducibility and the repeatability were $\sigma_{\bt} = \SI{0.28}{nT^2}$ and $\tau_{\bt} = \SI{0.02}{nT^2}$.

The most important conclusion here is that the repeatability of the mapping is better than the field reproducibility.
It means that the mapping uncertainty is not dominated by the performance of the mapping measurement and analysis methods.
Indeed, although the degaussing procedure and the opening and closing of the shield is already very reproducible, it still dominates the field map precision.

Another relevant point to highlight is the comparison between the repeatability and the propagated error calculated with the analysis method.
The values of these quantities are listed in Table~\ref{tab:comparisonErrors} for $\G$ and $\bt$.
\begin{table}
    \centering
    \caption{Reproducibility, repeatability and fit error of $\G$ and $\bt$ calculated from a global analysis of the $B_0$ maps. Reproducibility and repeatability are calculated with formulae~\ref{Eq:reproducibility} and \ref{Eq:repeatability} respectively.}
    \label{tab:comparisonErrors}
    \def\arraystretch{1.1}
    \setlength\tabcolsep{8pt}
    \begin{tabular}{cc|ccc}
    $x$ & Unit
    & $\sigma_x$  
    & $\tau_x$ 
    & $\Delta x_{\rm fit}$\\
    \hline \hline
    $\G$ & pT/cm & 0.56 & 0.38 & 0.19 \\
    $\bt$ & nT$^2$ & 0.28 & 0.02 & 0.02 \\
    \hline
    \end{tabular}
\end{table}
On one hand, the repeatability quantifies all the uncertainties due to measurement differences from one map recording to another, for example position errors or varying misalignements, or in particular drifts of the fluxgate offset. 
On the other hand, the fit error $\Delta x_{\rm fit}$ also takes several other error sources into account, the obvious one being a potential model incompleteness, since we only consider field modes $G_{l,m}$ with $l \leq 7$. As an example, some complex fields caused by a local contamination or a deformation on the magnetic shield could be impossible to describe with the limited set of coefficients we restrict ourselves to.
One might think that the fit error (propagated from the uncertainties on each Fourier coefficient $a_{m,\{ \rho,\phi,z \} ,i}$) should be bigger than or at least equal to the repeatability.
Nevertheless, it is not the case for $\G$, $\bt$ and for most of the generalized gradients, as correlations occur as the fluxgate drifts are slow.
While each ring individually fits well suggesting a lower uncertainty, considering the map as a whole the drift grows large.
We rescaled this fit error with the square root of the reduced fit $\chi^2$ to allow us to take the error due to the fluxgate drifts into account.
We therefore use the repeatability (rather than the error propagated from the fit) as our key metric of the measurement uncertainty for parameters extracted from a single map.

The global analysis of all $B_0$ maps was also used to compare the measurements with the simulations which is discussed in Sec.~\ref{Sec:simuResultsComparison}.
In the following section, we discuss the method to extract the value of the phantom gradient $\G$ and the transverse inhomogeneity $\bt$ for each nEDM sequence.
\FloatBarrier

\subsection{Gradient reconstruction method\label{Sec:linearity}}
We identified two possible methods to obtain the gradients from the mapping for each magnetic configuration corresponding to an nEDM datataking sequence.
The first option is to map all the different configurations used for EDM measurements and extract the gradients from the analysis of each individual map.
The second method is to use the linear dependence of the field on the applied coil currents and combine the analysis results of $B_0$ maps, trimcoil maps and guiding coil maps to reconstruct the magnetic field. 
Once we obtain the gradients with one of these methods, the calculation of the transverse inhomogeneity $\bt$ is simply an application of the formulae given in Appendix~\ref{App:Bt2Expression}.
In this section, we will briefly describe the global analysis of the coil maps, verify the linearity to validate the second method and then compare the accuracy of both methods.

Unlike for the $B_0$ coil, the currents used in the trimcoils during the EDM sequences changed from one magnetic configuration to another.
Therefore, to obtain the contribution to the gradients of each coil, the relation between the current flowing through the coil and the field produced had to be used.
This relation is linear in the case where no ferromagnetic material is present. 
In our case, the $B_0$ coil was within a large mu-metal shield which was responsible for 40\% of the $B_0$ field.
However, as the shield was far from the saturated regime, the field produced should have been linear in the coil currents. As we will show below our analysis proves that the linearity assumption was valid.

For every coil (trimcoils and guiding coils), one to five maps were taken with the coil powered with a current of 10 or \SI{20}{mA}.
Each time a coil map was taken, a map of the remnant field $B_{\rm rem}$ was recorded, too.
Both maps were analyzed and the gradients were extracted with the method described in Sec.~\ref{Sec:SingleMapAnalysis}.
The remnant field gradients were subtracted from the coil ones so that we consider only the field created by the coil.
The gradients thus obtained were then scaled with the value of the current flowing through the coil in order to obtain the gradients produced by \SI{1}{\micro A}.
Finally, when there were several maps of one coil, we combined them after analyzing them all by calculating the weighted mean.

With these coefficients and the results of the $B_0$ maps analysis, we are able to calculate the gradients of any magnetic-field configuration by using the linearity of the gradients,
\begin{equation}
\label{Eq:linearity}
\G = \G_{B_0}^{\uparrow {\rm or} \downarrow} + \sum_c^{N_{\rm coils}} i_c \widehat{g}_{c},
\end{equation}
where $\G_{B_0}^{\uparrow {\rm or} \downarrow}$ is the average value of $\G$ measured in up or down $B_0$ maps, estimated with Eq.~\ref{Eq:globalAverage}, $N_{\rm coils}$ is the number of additional coils, $i_c$ is the current 
and $\widehat{g}_{c}$ is the gradient produced by \SI{1}{\micro A} in coil $c$.

\begin{figure}[h!]
\centering
    \subfigure[]{
        \includegraphics[width = \columnwidth]{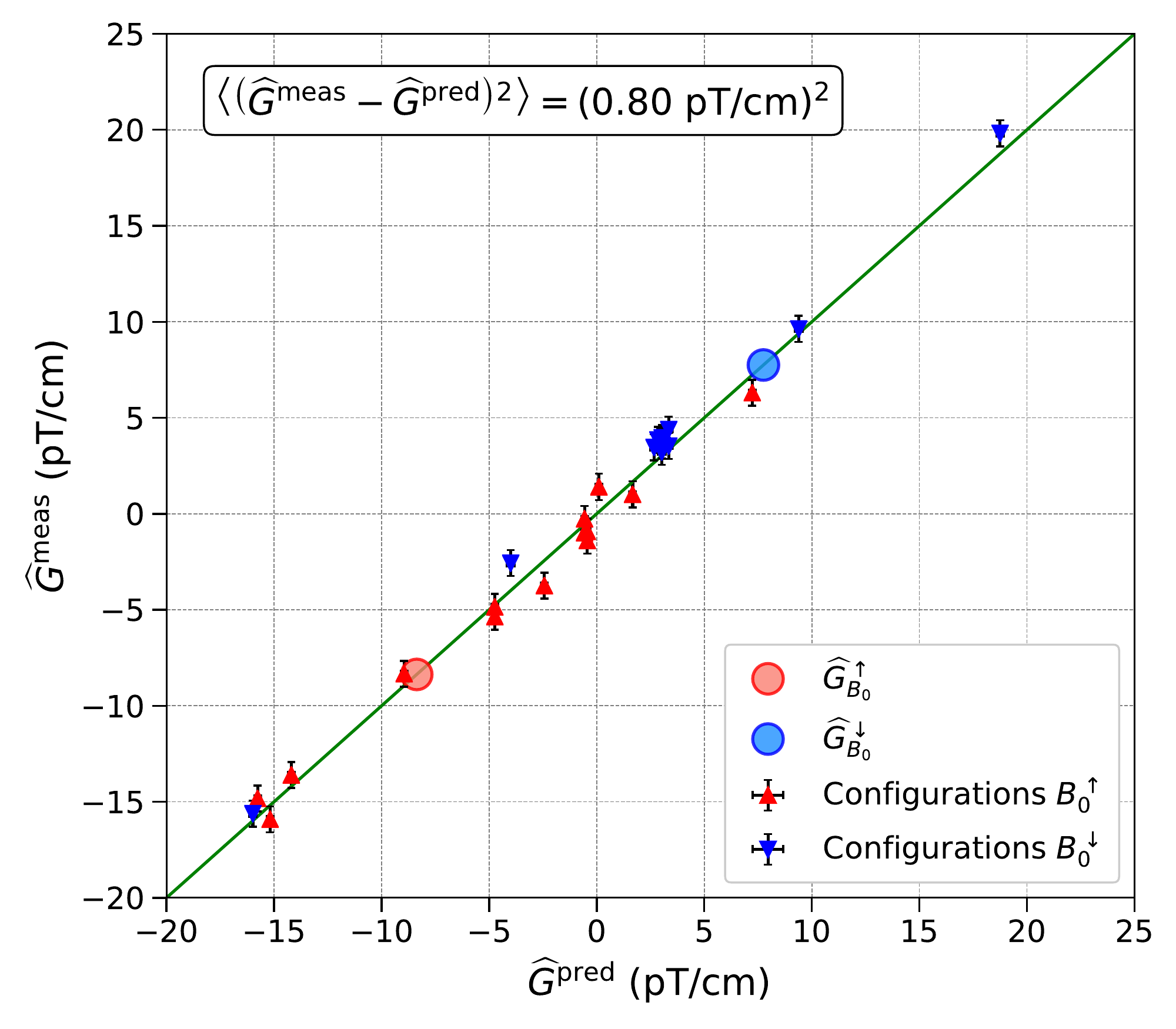}
        \label{Fig:linearity1}}
    \hfill
    \subfigure[]{\includegraphics[width = \columnwidth]{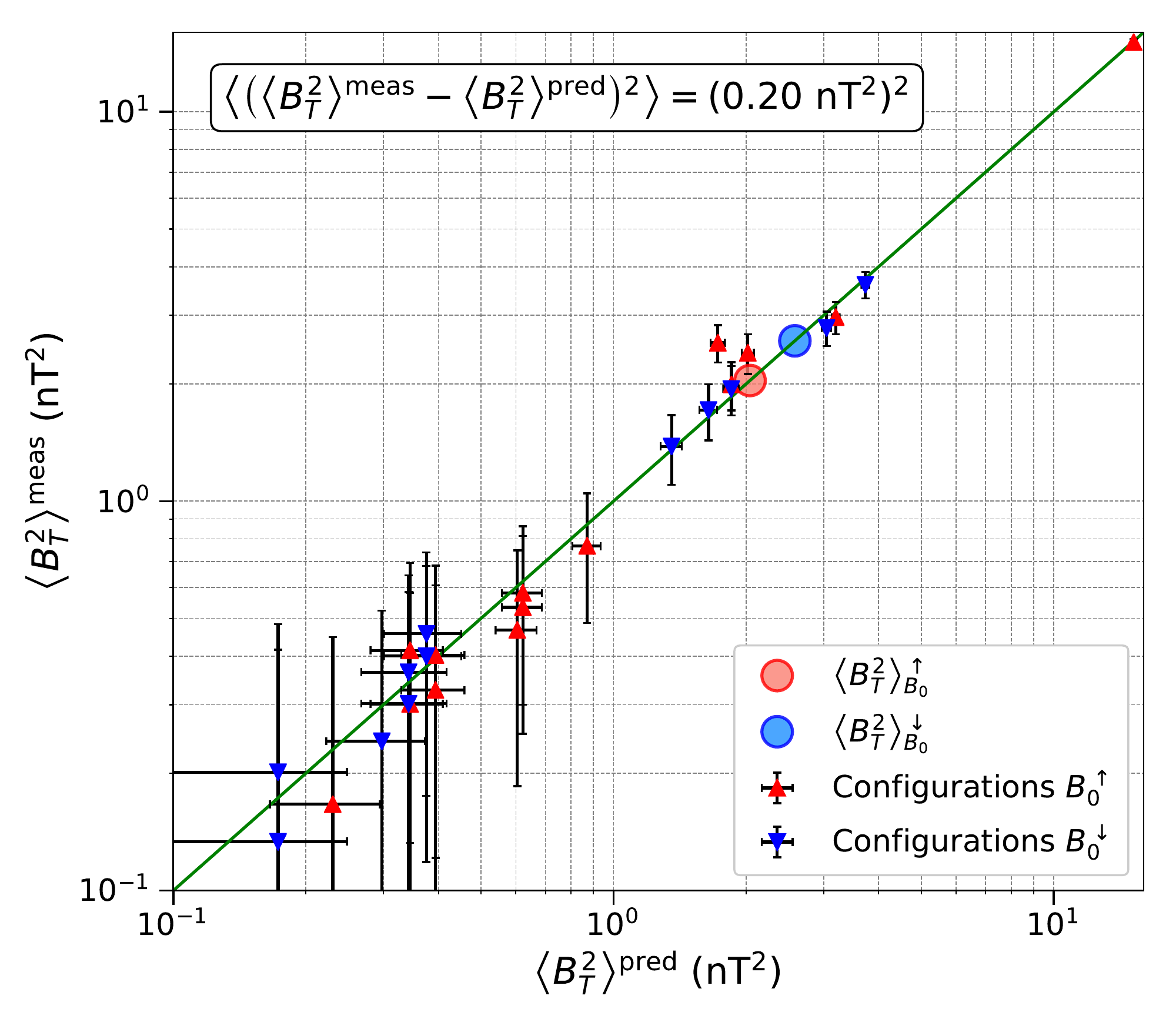}\label{Fig:linearity2}}
    \caption{Comparison of the measured and predicted values for the maps of the nEDM sequence configurations.
    The green line is the first bisector $y=x$.
     The RMS written in the top left corner of each plot is the mean square difference square root. 
    (a) Comparison for the gradient $\G$. The large dots are the average values of the gradient extracted from the analysis of the $B_0$ maps, see \autoref{Fig:reproducibility}.
    (b) Comparison for the transverse inhomogeneity $\langle B_T^2 \rangle$. The $\langle B_T^2 \rangle \sim \SI{15}{nT^2}$ point in the upper right corner corresponds to the magnetic configuration of one of the first nEDM data sequences, when the uniformity optimisation method~\cite{Abel2020} was not used yet.
    \label{Fig:linearity}}
\end{figure}

In order to check the validity of this prediction method, we compared the gradients extracted from the maps of the EDM sequence configurations to their predicted values using the linear superposition method.
The results of this comparison for the gradient $\G$ and for the transverse inhomogeneity $\langle B^2_{\rm T} \rangle$ are shown on Figs.~\ref{Fig:linearity1} and~\ref{Fig:linearity2}.
For both $\G$ and $\langle B^2_{\rm T} \rangle$, one can see that the prediction and the measurement are in good agreement.
We can therefore validate the accuracy of the prediction method, since it reliably reconstructs the measured gradients.
The mean square differences of the comparison are:
\begin{eqnarray}
    &\left\langle \left( \G^{\rm meas} - \G^{\rm pred} \right)^2 \right\rangle = (0.80 \ {\rm pT/cm})^2 \\
    &\left\langle \left( {\bt}^{\rm meas} - {\bt}^{\rm pred} \right)^2 \right\rangle = (0.20 \ {\rm nT^2})^2 .
\end{eqnarray}
There are several contributions to these differences.
The main contribution for both $\G$ and $\bt$ is the $B_0$ reproducibility (\SI{0.56}{pT/cm} for $\G$ and $\SI{0.28}{nT^2}$ for $\bt$).
On the one hand, for the transverse inhomogeneity $\bt$, the mean square difference is a little smaller than the reproducibility.  
On the other hand, we can see for the phantom gradient $\G$ that other sources of error seem to contribute.
One of them is the error arising from the incorporation of the trimcoil and guiding coil contributions to the prediction.
To estimate the size of this error, we did another specific comparison to eliminate the $B_0$ reproducibility contribution.
We compared the gradients of the sequence maps subtracted from the gradients of $B_0$ maps taken in the same group of measurements (no shield degaussing) with the prediction coming from the additional coils. 
For $\G$, the mean square difference of this second comparison was $(\SI{0.70}{pT/cm})^2$.
The quadratic contributions to this difference are:
\begin{itemize}
    \item The mapping method uncertainty, for which we take the repeatability $\tau_{\G}$. It must be taken into account twice, once for the sequence map and once for the $B_0$ map: $2\!\cdot\!(\SI{0.38}{pT/cm})^2$.
    \item The coils prediction error, which can be deduced from the other contribution: $(\SI{0.45}{pT/cm})^2$. 
\end{itemize}
One can see that the coils prediction error is the same order of magnitude as the repeatability. 
However, it is still subdominant compared to the field reproducibility, which remains the limiting uncertainty.
We now have a full explanation of all contributions to the uncertainties and can compare the accuracy of both methods to obtain the gradients for one magnetic configuration.

As said in the beginning of this section, the two methods to obtain the gradients for one nEDM sequence magnetic configuration are: 
\begin{enumerate}
    \item Extracting them by offline measurement of the same magnetic-field configuration.
    \item Calculating them by combining individual offline measurements of all the coils, $B_0$ and all trim coils, contributing to the generation of the field.
\end{enumerate}
Since the largest systematic effect on the EDM result is due to the gradient $\G$, we will compare the uncertainties for this gradient to determine which method is more accurate.
However, for each individual gradient $G_{l,m}$, the uncertainty sources are the same, so the uncertainty expressions are identical in form.
The expressions of the uncertainty are, for the first method,
\begin{equation}
\label{Eq:configMap}
\left( \Delta\G^{\rm meas(1)} \right)^2 = \sigma_{\G}^2 + \sigma_{\G}^2 + \tau_{\G}^2,
\end{equation}
and for the second method,
\begin{equation}
\label{Eq:prediction}
\left( \Delta\G^{\rm pred(2)} \right)^2 = \sigma_{\G}^2 + \frac{\sigma_{\G}^2  + \tau_{\G}^2}{N^{\uparrow {\rm or} \downarrow}} + \sum_c \left( i_c \Delta\G_c \right)^2.
\end{equation}
%

As the shield is opened and degaussed between neutron datataking and mapping measurements, the field reproducibility is the largest contribution to the prediction error, and is unavoidable. This is the first term, and the same in each expression.

For the first method, the second contribution to Eq.~\ref{Eq:configMap} is again the $B_0$ field reproducibility.
Since this method uses the analysis of one map, 
the reproducibility error has to be taken into account again.
The last term is then simply the uncertainty coming from the mapping analysis of one map: the mapping repeatability $\tau_{\G}$.
With the second method, the other contributions to the uncertainty in Eq.~\ref{Eq:prediction} are the errors on the prediction accuracy.
The two last terms are the respective uncertainties of the terms of Eq.~\ref{Eq:linearity}.

The $B_0$ field reproducibility is the main contribution among all these terms.
Therefore, one can see from the expressions in Eq.~\ref{Eq:configMap} and \ref{Eq:prediction} that if all other contributions are negligible, the uncertainty coming from the first method $\Delta\G^{\rm meas(1)}$ is bigger than the one from the second method, $\Delta\G^{\rm pred(2)}$, by a factor $\sqrt{2}$.
It turns out that the other terms are in fact not negligible but $\Delta\G^{\rm meas(1)}$ is still bigger than $\Delta\G^{\rm pred(2)}$. 
We therefore chose the second method to predict the gradients of all the nEDM sequence magnetic configurations. 
This has the additional benefit that any anomalous maps would be easily identified and removed from the analysis.
Indeed, with the second method, all $B_0$ and coil maps were measured multiple times.
Contrastingly, most of the 22 nEDM data sequence base configurations (as defined in \cite{Abel2020_2}, optimised field configurations used for datataking that were modified only by adding small well characterised vertical gradients up to around $\abs{\Delta G_{1,0}} \leq \SI{30}{pT/cm}$) were mapped only once.
\FloatBarrier

\section{Comparison with simulations\label{Sec:simuResultsComparison}}
As said in Sec.~\ref{Sec:ReprodAndRepeat}, the global analysis method of the $B_0$ maps can be applied to compare the results of the measurements with the simulations.
The values of the gradients for the allowed modes, their measurement uncertainties and a relative difference with the simulation are listed in Table~\ref{tab:allowedModeComparison}.
\begin{table}
    \caption{
    Ansys simulation predicted value for the magnetic-field modes allowed by the symmetries of the $B_0$ coil and comparison with the measured values. 
    The value $\Delta G_{l,m}^{B_0 \ \rm pred}$ here corresponds to the error on the prediction of the gradient produced by $B_0$ when in up configuration and is $\Delta G_{l,m}^{B_0 \ \rm pred} = ((\sigma_{G_{l,m}}^2  + \tau_{G_{l,m}}^2)/N^{\uparrow})^{1/2}$.
    \label{tab:allowedModeComparison}}
    \setlength\tabcolsep{8pt}
    \begin{tabular}{c|ccc}
        Mode 
        & $G_{l,m}^{\rm meas}$ 
        & $\Delta G_{l,m}^{B_0 \ \rm pred}$  
        & $\frac{\left|G_{l,m}^{\rm simu} - G_{l,m}^{\rm meas}\right|}{G_{l,m}^{\rm meas}}$ \\
        Unit & (pT/cm$^{l}$) & (pT/cm$^{l}$) & -- \\
        \hline \hline
        $G_{0,0}$ & $1034.15 \times 10^3$ & $0.23 \times 10^3$ & 0.03\% \\
        $G_{2,0}$ & $-7.62$ & $0.06$ & 21.46\% \\
        $G_{2,2}$ & $2.24$ & $0.02$ & 47.55\% \\
        $G_{4,0}$ & $-4.03 \times 10^{-3}$ & $0.09 \times 10^{-3}$ & 9.97\% \\
        $G_{4,2}$ & $1.59 \times 10^{-3}$ & $0.01 \times 10^{-3}$ & 13.67\% \\
        $G_{4,4}$ & $-1.10 \times 10^{-4}$ & $0.03 \times 10^{-4}$ & 21.13\% \\
        $G_{6,0}$ & $-1.35 \times 10^{-6}$ & $0.05 \times 10^{-6}$ & 13.48\% \\
        $G_{6,2}$ & $2.57 \times 10^{-7}$ & $0.04 \times 10^{-7}$ & 8.07\% \\
        $G_{6,4}$ & $-1.03 \times 10^{-7}$ & $0.02 \times 10^{-7}$ & 23.09\% \\
        $G_{6,6}$ & $-1.49 \times 10^{-8}$ & $0.12 \times 10^{-8}$ & 166.22\% \\
        \hline
    \end{tabular}
\end{table}
These measured gradients can be compared with the ones simulated, in Table~\ref{tab:allowedMode}.
One can see that the uniform mode $G_{0,0}$ is very well predicted (0.03\%) by the simulations.
The other allowed modes are predicted within 20\% of agreement with the measurement, except for the $G_{2,2}$ and $G_{6,6}$ modes.
For this last mode, it can be explained by the precision of the analysis method.
Indeed, since the analysis is performed up to order $l=6$ and $m=6$, the order $G_{6,6}$ is less constrained in the harmonic fit step of the analysis and is also influenced by higher order components that are not fitted separately.
Concerning the other modes, for both the simulation and the measurement, the uncertainties cannot explain the differences.
By changing the parameters of the simulation, its numerical precision can be estimated, and this also does not provide an explanation. 
We therefore assume that the difference is due to the simplification of the system geometry (perfectly symmetric coil and shield, small shield holes ignored, etc.).
However, what is to remember is that we are able to predict very accurately the uniform term for a field produced by a coil in a multiple-layer shield and obtain the magnitude of the higher order allowed modes of the field. 
\FloatBarrier

\section{Discussion}

\subsection{EDM corrections\label{Sec:edmCorr}}
In this section we discuss how the magnetic corrections affect the analysis and result of the nEDM measurement.
In total 99 nEDM measurement sequences were used in the analysis.
For each of these sequences, we correct the measured ratio $\R$ with $\bt$ and the measured EDM $d_n$ with the phantom $\G$, using Equations \ref{eqn:quadShift} and \ref{falseEDM_Ghat}, respectively.
Then, all these sequences are analysed together, and the apparent nEDM $d_n^{\rm corr}$ and $\R^{\rm corr}$ found in each sequence are fit to Equation \ref{eqn:crossingLines} to account for the gravitational shift $\delta_{\mathrm{grav}}$ and the fraction of the mercury induced false EDM proportional to $G_\mathrm{grav}$. 
Since for a fixed $G_\mathrm{grav}$ the sign of $\delta_\mathrm{grav}$ inverts while the sign of $d^{\rm false}_{n \leftarrow {\rm Hg}}$ does not, this fit can be visualized as fitting a pair of lines of opposite, fixed, slope (see Fig.~4 in \cite{Abel2020}). 
Where the two lines cross, it can be inferred that $G_\mathrm{grav} = 0$, and so these two effects are eliminated. As such, this step is sometimes referred to as the ``crossing lines'' or ``crossing point'' analysis.

As detailed in Sec.~\ref{sec:CorrStrat}, the corrections affect the ratio $\R$. Therefore, the corrections coming from the transverse inhomogeneity $\langle B_{\rm T}^2\rangle$ shift the crossing point nEDM value if they are different for each polarity of the $B_0$ field.
If these shifts are the same for both signs of $B_0$, then the crossing point $\R$ will be affected, but the crossing point $d_n$ will not be affected.
In each of the sequences, a correction between $2\times10^{-7}$ and $175\times10^{-7}$ was subtracted from the measured ratio $\mathcal{R}$.
After this procedure, the crossing point was shifted by $\left(0 \pm 5\right) \times 10^{-28}\, e \, {\rm cm}$, where the uncertainty given reflects the overall systematic uncertainty from the correction of the shift due to $\langle B_{\rm T}^2\rangle$.
The correction of $\langle B_{\rm T}^2\rangle$ thus did not impact the value of the measured nEDM\@.
However, it marginally improved the quality of the crossing point fit, corresponding to a reduction in $\chi^2$ of 4\%.

\begin{figure}
    \includegraphics[width = \columnwidth]{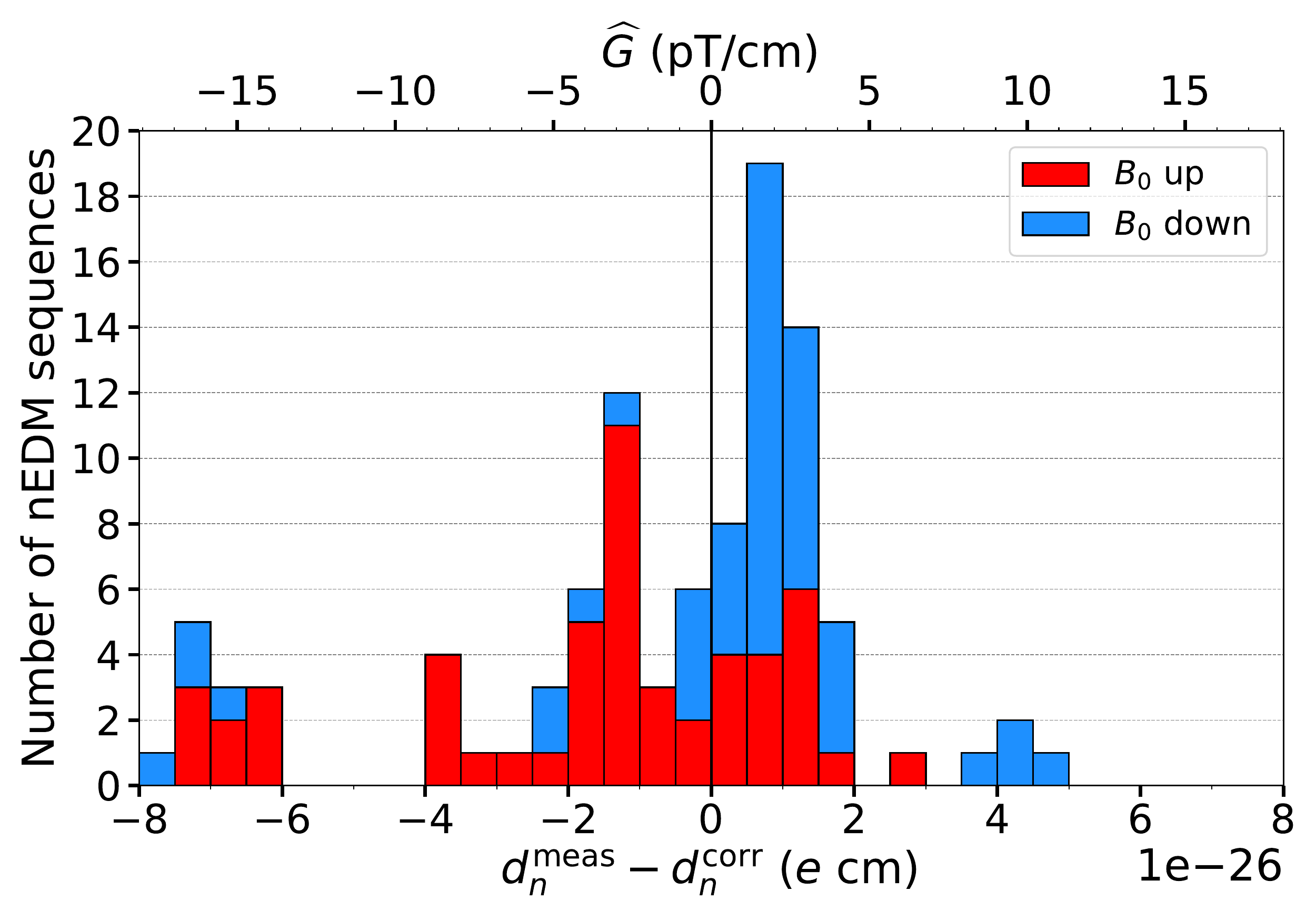}
    \caption{Predicted values of $\G$ and the corresponding corrections of $d_n$ for the 99 nEDM measurement sequences. 
    \label{Fig:nedmCorr}}
\end{figure}

The values of the magnetic-field related corrections of $d_n$ coming from the predicted gradient $\G$ for the 99 sequences can be seen in \autoref{Fig:nedmCorr}.
One can see that the values of $\G$ for the sequences are different from the ones produced by the $B_0$ coil alone.
Since we used the trimcoils to compensate small inhomogeneities in the $B_0$ field (using the optimisation technique described in \cite{Abel2020} after each degaussing) and also to produce a particular value of the gradient $G_{\rm grav}$ for each measurement sequence, a unique value of $\G$ was calculated for each sequence. 
The values of the $\G$ corrections on some sequences can reach up to seven times the global statistical uncertainty of the EDM\@.
Once we took all $\G$ corrections into account, the shift of the crossing point value was $\left( 69 \pm 10\right) \times 10^{-28}\, e \, {\rm cm}$. 
This shift of the nEDM measurement is about 60\% of the nEDM statistical error and is the largest systematic effect.
The uncertainty from that effect is the biggest source of systematic error in \cite{Abel2020_2}.

\subsection{Conclusion}
We discussed the offline measurement of the magnetic-field non-uniformity for the most sensitive neutron EDM measurement\,\cite{Abel2020_2} and compared two methods for a calculation of mandatory systematic corrections (see Eq.\ref{Eq:linearity} and \ref{Eq:prediction}). 
As explained in Sec.~\ref{Sec:SystematicEffects}, the predicted values of the gradient $\G$ and the transverse inhomogeneity $\bt$ are needed to correct the values of $d_n$ and $\mathcal{R}$ for the crossing point method.
The explanation of this method and its result can be found in~\cite{Abel2020_2}.

This paper concludes the trilogy of articles~\cite{Abel2019,Abel2020} describing the effects, control and correction of magnetic-field non-uniformity in a neutron EDM measurement experiment.
The experience gained, the knowledge acquired, and the techniques developed during experiments using the single chamber nEDM will be extremely valuable for future experiments, such as the n2EDM experiment at PSI \cite{Ayres2021TDR}.

\begin{acknowledgements}
The experimental data were taken at PSI Villigen. 
We acknowledge the excellent support provided by the PSI technical groups and by various services of the collaborating universities and research laboratories. 
The authors would like to thank their collaborators from the LPC Caen CAD group and workshop for their deep involvement in the design, manufacture and assembly of the mapper.
We  gratefully  acknowledge  financial  support  from the  Swiss  National  Science  Foundation  through  projects 137664 (PSI), 117696 (PSI), 144473 (PSI), 126562 (PSI), 181996
(Bern), 200441 (ETH), 172639 (ETH) and 140421 (Fribourg); and from STFC, via grants ST/M003426/1, ST/N504452/1 and ST/N000307/1. The LPC Caen and the LPSC  Grenoble  acknowledge  the  support  of  the  French Agence  Nationale  de  la  Recherche  (ANR)  under  reference ANR-09-BLAN-0046 and the ERC project 716651-NEDM. The Polish collaborators wish to acknowledge support from the National Science Center, Poland, under grants 2016/23/D/ST2/00715, 2018/30/M/ST2/00319 and 2020/37/B/ST2/02349. P.~Mohanmurthy  acknowledges grant SERI-FCS 2015.0594.  This work was also partly supported by the Fund for Scientific Research Flanders (FWO), and Project GOA/2010/10 of the KU~Leuven.  In addition we are grateful for access granted to the computing grid infrastructure PL-Grid.

\end{acknowledgements}

\appendix

\section{Harmonic polynomials in cylindrical coordinates\label{appendix_polyCynlindrical}}
It is useful to derive the expressions of the harmonic modes in cylindrical coordinates 
$(\rho, \phi, z)$ since this coordinate system is the most relevant for the mapping analysis. 
The polynomials can be obtained by deriving the formula of the magnetic potential cited in \cite{Abel2019}:
\begin{equation}
\label{magneticPotential}
\Sigma_{l,m} = C_{l,m}(\phi) r^l P_l^{|m|}(\ct),
\end{equation}
where $P_l^m$ are the associated Legendre polynomials and
\begin{eqnarray}
C_{l,m}(\phi) & = & \frac{(l-1)! (-2)^{|m|}}{(l+|m|)!} \cos(m \phi) \quad {\rm for } \quad m \geq 0 \\
\nonumber
C_{l,m}(\phi) & = & \frac{(l-1)! (-2)^{|m|}}{(l+|m|)!} \sin(|m| \phi) \quad {\rm for } \quad m < 0.
\end{eqnarray}
The radial, azimuthal and vertical components respectively of the mode $l,m$ are then given by 
\begin{eqnarray}
\Pi_{\rho, l, m} & = & \partial_\rho \Sigma_{l+1, m} \\
\Pi_{\phi, l, m} & = &  \frac{1}{\rho} \partial_\phi \Sigma_{l+1, m} \\
\Pi_{z, l, m} & = & \partial_z \Sigma_{l+1, m}, 
\end{eqnarray}
and are listed up to order 7 in Tables~\ref{adequateCylindrical}, \ref{adequateCylindrical2} and~\ref{adequateCylindrical3}.
\begin{table*}
\caption{
The basis of harmonic polynomials sorted by order in cylindrical coordinates, to order $l=0$ to $l=4$. 
\label{adequateCylindrical}}
\begin{tabular}{cc|c@{\hspace{1cm}}c@{\hspace{1cm}}c}
$l$ & $m$ & $\Pi_\rho$ & $\Pi_\phi$  & $\Pi_z$ \\
\hline \hline
$0$ & $-1$  & $\sin \phi$         & $\cos \phi$         & $0$          \\
$0$ & $0$  & $0$                 & $0$                 & $1$          \\
$0$ & $1$  & $\cos \phi$         & $-\sin \phi$        & $0$          \\
\hline
$1$ & $-2$  & $\rho \sin 2\phi$   & $\rho \cos 2\phi$   & $0$               \\
$1$ & $-1$  & $z \sin \phi$       & $z \cos \phi$       & $\rho \sin \phi$  \\
$1$ & $0$  & $-\frac12 \rho$     & $0$                 & $z$               \\
$1$ & $1$  & $z \cos \phi$       & $-z \sin \phi$      & $\rho \cos \phi$  \\
$1$ & $2$  & $\rho \cos 2\phi$   & $- \rho \sin 2\phi$ & $0$               \\
\hline
$2$ & $-3$ & $\rho^2 \sin 3\phi$  & $\rho^2 \cos 3\phi$ & $0$          \\
$2$ & $-2$ & $2 \rho z \sin 2\phi$     & $2\rho z \cos 2\phi$        & $\rho^2 \sin 2\phi$   \\
$2$ & $-1$ & $\frac14 (4z^2-3\rho^2)\sin \phi$ & $\frac14 (4z^2-\rho^2)\cos \phi$ & $2\rho z \sin \phi$ \\
$2$ & $0$ & $- \rho z$           & $0$                 & $-\frac12 \rho^2 + z^2$      \\
$2$ & $1$ & $\frac14 (4z^2-3\rho^2)\cos \phi$ & $\frac14 (\rho^2-4z^2)\sin \phi$ & $2 \rho z \cos \phi$ \\
$2$ & $2$ & $2 \rho z \cos 2\phi$    & $-2 \rho z\sin 2\phi$     & $\rho^2 \cos 2\phi$    \\
$2$ & $3$ & $\rho^2 \cos 3\phi$ & $-\rho^2 \sin 3\phi$ & $0$         \\
\hline
$3$ & $-4$ & $\rho^3 \sin 4\phi$      & $\rho^3 \cos 4\phi$        & $0$            \\
$3$ & $-3$ & $3 \rho^2 z \sin 3\phi$ & $3 \rho^2 z \cos 3\phi$   & $\rho^3 \sin 3\phi$    \\
$3$ & $-2$ & $\rho (3z^2-\rho^2) \sin 2\phi$ & $\frac12 \rho (6z^2-\rho^2) \cos 2\phi$ & $3 \rho^2 z \sin 2\phi$     \\
$3$ & $-1$ & $\frac 14z(4z^2-9\rho^2)\sin \phi$ & $\frac 14z(4z^2-3\rho^2) \cos \phi$   & $\rho(3 z^2- \frac 34 \rho^2) \sin \phi$    \\
$3$ & $0$ & $\frac38\rho(\rho^2-4z^2)$  & $0$  & $\frac12z(2z^2-3\rho^2)$  \\
$3$ & $1$ & $\frac 14z(4z^2-9\rho^2)\cos \phi$ & $\frac 14z(3\rho^2-4z^2)\sin \phi$ & $\rho(3z^2 - \frac 34 \rho^2)\cos \phi$ \\
$3$ & $2$ & $\rho(3z^2-\rho^2)\cos 2\phi$    & $\frac12\rho(\rho^2-6z^2) \sin 2\phi$   & $3 \rho^2 z \cos 2\phi$  \\
$3$ & $3$ & $3 \rho^2 z \cos 3\phi$  & $-3 \rho^2 z \sin 3\phi$   & $\rho^3 \cos 3\phi$   \\
$3$ & $4$ & $\rho^3 \cos 4\phi$      & $- \rho^3 \sin 4\phi$      & $0$            \\
\hline
$4$ & $-5$  &$\rp{4} \sph{5}$  &$\rp{4} \cph{5}$  &$0$  \\
$4$ & $-4$  &$4\rp{3} z \sph{4}$  &$4\rp{3} z \cph{4}$  &$\rp{4} \sph{4}$  \\
$4$ & $-3$  &$\frac{1}{4}(24\rp{2} \zp{2}-5\rp{4}) \sph{3}$  &$\frac{3}{4}(8\rp{2} \zp{2}-\rp{4}) \cph{3}$  &$4\rp{3} z \sph{3}$  \\
$4$ & $-2$  &$4(\rho\zp{3}-\rp{3}z) \sph{2}$   &$2(2\rho\zp{3}-\rp{3}z) \cph{2}$  &$(6\rp{2}\zp{2}-\rp{4}) \sph{2}$ \\
$4$ & $-1$  &$\frac{1}{8}(8\zp{4}-36\rp{2} \zp{2}+5\rp{4}) \sphi$  &$\frac{1}{8}(8\zp{4}-12\rp{2} \zp{2}+\rp{4}) \cphi$  &$(4\rho\zp{3}- 3\rp{3}z) \sphi$  \\
$4$ &  $0$  &$\frac{1}{2}(3\rp{3} z-4\rho \zp{3})$  &$0$  &$\frac{1}{8}(8\zp{4}-24\rp{2} \zp{2}+3\rp{4})$ \\
$4$ &  $1$  &$\frac{1}{8}(8\zp{4}-36\rp{2} \zp{2}+5\rp{4}) \cphi$ &$-\frac{1}{8}(8\zp{4}-12\rp{2} \zp{2}+\rp{4}) \sphi$ &$(4\rho\zp{3}- 3\rp{3}z) \cphi$  \\
$4$ &  $2$  &$4(\rho\zp{3}-\rp{3}z) \cph{2}$  &$-2(2\rho\zp{3}-\rp{3}z) \sph{2}$  &$(6\rp{2}\zp{2}-\rp{4}) \cph{2}$ \\
$4$ &  $3$  &$\frac{1}{4}(24\rp{2} \zp{2}-5\rp{4}) \cph{3}$ &$-\frac{3}{4}(8\rp{2} \zp{2}-\rp{4}) \sph{3}$  &$4\rp{3} z \cph{3}$ \\
$4$ &  $4$  &$4\rp{3} z \cph{4}$  &$-4\rp{3} z \sph{4}$  &$\rp{4} \cph{4}$  \\
$4$ &  $5$  &$\rp{4} \cph{5}$  &$-\rp{4} \sph{5}$  &$0$  \\
\hline
\end{tabular}
\end{table*}
\begin{table*}
\caption{
The basis of harmonic polynomials sorted by order in cylindrical coordinates, from order $l=5$ to $l=6$. 
\label{adequateCylindrical2}}
\resizebox{\textwidth}{!}{
\begin{tabular}{cc|c@{\hspace{0.3cm}}c@{\hspace{0.3cm}}c}
$l$ & $m$ & $\Pi_\rho$ & $\Pi_\phi$  & $\Pi_z$ \\
\hline \hline
$5$ & $-6$  &$\rp{5} \sph{6}$  &$\rp{5} \cph{6}$  &$0$  \\
$5$ & $-5$  &$5\rp{4} z \sph{5}$  &$5\rp{4} z \cph{5}$  &$\rp{5} \sph{5}$  \\
$5$ & $-4$  &$\frac12(20\rp{3} \zp{2}-3\rp{5}) \sph{4}$  &$\rho^3 (10z^2 - \rho^2) \cph{4}$  &$5\rp{4} z \sph{4}$  \\
$5$ & $-3$  &$\frac54(8  \rp{2} \zp{3}-5\rp{4} z) \sph{3}$  &$\frac{5}{4}(8  \rp{2} \zp{3}-3\rp{4} z) \cph{3}$  &$\frac{5}{4}(8\rp{3} \zp{2}-\rp{5}) \sph{3}$  \\
$5$ & $-2$  &$\frac{5}{16}(16\rho \zp{4}-32\rp{3} \zp{2}+3\rp{5}) \sph{2}$  &$\frac{5}{16}(16\rho \zp{4}-16\rp{3} \zp{2}+\rp{5}) \cph{2}$  &$5(2  \rp{2} \zp{3}-\rp{4} z) \sph{2}$  \\
$5$ & $-1$  &$\frac18(8\zp{5}-60  \rp{2} \zp{3}+25\rp{4} z) \sphi$  &$\frac{1}{8}(8\zp{5}-20  \rp{2} \zp{3}+5\rp{4} z) \cphi$  &$\frac{5}{8}(8\rho \zp{4}-12\rp{3} \zp{2}+\rp{5}) \sphi$\\
$5$ &  $0$  &$\frac{5}{16}(-8\rho \zp{4}+12\rp{3} \zp{2}-\rp{5})$  &$0$  & $\frac{1}{8}(8\zp{5}-40  \rp{2} \zp{3}+15\rp{4} z)$  \\
$5$ &  $1$  &$\frac{1}{8}(8\zp{5}-60  \rp{2} \zp{3}+25\rp{4} z) \cphi$  &$-\frac{1}{8}(8\zp{5}-20 \rp{2} \zp{3}+5\rp{4} z) \sphi$  &$\frac{5}{8}(8\rho \zp{4}-12\rp{3} \zp{2}+\rp{5}) \cphi$\\
$5$ &  $2$  &$\frac{5}{16}(16\rho \zp{4}-32\rp{3} \zp{2}+3\rp{5}) \cph{2}$  &$-\frac{5}{16}(16\rho \zp{4}-16\rp{3} \zp{2}+\rp{5}) \sph{2}$  &$5(2  \rp{2} \zp{3}-\rp{4} z) \cph{2}$  \\
$5$ &  $3$  &$\frac{5}{4}(8  \rp{2} \zp{3}-5\rp{4} z) \cph{3}$  &$-\frac{5}{4}(8  \rp{2} \zp{3}-3\rp{4} z) \sph{3}$  &$\frac{5}{4}(8\rp{3} \zp{2}-\rp{5}) \cph{3}$  \\
$5$ &  $4$  &$\frac{1}{2}(20\rp{3} \zp{2}-3\rp{5}) \cph{4}$  &$-\rho^3 (10z^2 - \rho^2) \sph{4}$  &$5\rp{4}  \cph{4}z$  \\
$5$ &  $5$  &$5\rp{4} z \cph{5}$  &$-5\rp{4} z \sph{5}$  &$\rp{5} \cph{5}$  \\
$5$ &  $6$  &$\rp{5} \cph{6}$  &$-\rp{5} \sph{6}$  &$0$  \\
\hline
$6$ & $-7$  &$\rp{6} \sph{7}$  &$\rp{6} \cph{7}$  &$0$  \\
$6$ & $-6$  &$6\rp{5} z \sph{6}$  &$6\rp{5} z \cph{6}$  &$\rp{6} \sph{6}$  \\
$6$ & $-5$  &$\frac14 \rho^4 (60z^2 - 7\rho^2) \sph{5}$ &$\frac54 \rho^4 (12z^2 - \rho^2) \cph{5}$ &$6\rho^5z \sph{5}$ \\
$6$ & $-4$  &$\rho^3 z (20z^2 - 9\rho^2) \cph{4}$ &$2\rho^3z (10z^2 - 3\rho^2) \cph{4}$ &$\frac32 \rho^4 (10z^2 - \rho^2) \sph{4}$ \\
$6$ & $-3$  &$\frac{3}{16} \rho^2 (80z^4 - 100\rho^2 z^2 + 7\rho^4) \cph3$ &$\frac{3}{16}\rho^2 (80z^4 - 60\rho^2z^2 + 3\rho^4) \cph{3}$ &$\frac52 \rho^3z (8z^2 - 3\rho^2) \sph{3}$ \\
$6$ & $-2$  &$\frac18 \rho z (48z^4 - 160\rho^2 z^2 + 45\rho^4) \cph{2}$ &$\frac18 \rho z (48z^4 - 80\rho^2z^2 + 15\rho^4) \cph{2}$ &$\frac{15}{16} \rho^2 (16z^4 - 16\rho^2z^2 + \rho^4) \sph{2}$ \\
$6$ & $-1$  &$\frac{1}{64} (64z^6 - 720\rho^2 z^4 + 600\rho^4 z^2 - 35\rho^6) \cphi$ &$\frac{1}{64} (64z^6 - 240\rho^2z^4 + 120\rho^4z^2 - 5\rho^6) \cphi$ &$\frac34 \rho z (8z^4 - 20\rho^2z^2 + 5\rho^4) \sphi$ \\
$6$ &  $0$  &$\frac{3}{8} \rho (-8z^5 + 20\rho^2 z^3 - 5\rho^4 z)$ &$0$ &$\frac{1}{16} (16z^6 - 120\rho^2z^4 + 90\rho^4z^2 - 5\rho^6)$ \\
$6$ &  $1$  &$\frac{1}{64} (64z^6 - 720\rho^2 z^4 + 600\rho^4 z^2 - 35\rho^6) \sphi$ &$-\frac{1}{64} (64z^6 - 240\rho^2z^4 + 120\rho^4z^2 - 5\rho^6) \sphi$ &$\frac34 \rho z (8z^4 - 20\rho^2z^2 + 5\rho^4) \cphi$ \\
$6$ &  $2$  &$\frac18 \rho z (48z^4 - 160\rho^2 z^2 + 45\rho^4) \sph{2}$ &$-\frac18 \rho z (48z^4 - 80\rho^2z^2 + 15\rho^4) \sph{2}$ &$\frac{15}{16} \rho^2 (16z^4 - 16\rho^2z^2 + \rho^4) \cph{2}$ \\
$6$ &  $3$  &$\frac{3}{16} \rho^2 (80z^4 - 100\rho^2 z^2 + 7\rho^4) \sph3$ &$-\frac{3}{16}\rho^2 (80z^4 - 60\rho^2z^2 + 3\rho^4) \sph{3}$ &$\frac52 \rho^3z (8z^2 - 3\rho^2) \cph{3}$ \\
$6$ &  $4$  &$\rho^3 z (20z^2 - 9\rho^2) \sph{4}$ &$-2\rho^3z (10z^2 - 3\rho^2) \sph{4}$ &$\frac32 \rho^4 (10z^2 - \rho^2) \cph{4}$ \\
$6$ &  $5$  &$\frac14 \rho^4 (60z^2 - 7\rho^2) \cph{5}$ &$-\frac54 \rho^4 (12z^2 - \rho^2) \sph{5}$ &$6\rho^5z \cph{5}$ \\
$6$ &  $6$  &$6\rp{5} z \cph{6}$  &$-6\rp{5} z \sph{6}$  &$\rp{6} \cph{6}$  \\
$6$ &  $7$  &$\rp{6} \cph{7}$  &$-\rp{6} \sph{7}$  &$0$  \\
\hline
\end{tabular}}
\end{table*}
\begin{table*}
\caption{
The basis of harmonic polynomials sorted by order in cylindrical coordinates of order $l=7$. 
\label{adequateCylindrical3}}
\resizebox{\textwidth}{!}{
\begin{tabular}{cc|c@{\hspace{0.3cm}}c@{\hspace{0.3cm}}c}
$l$ & $m$ & $\Pi_\rho$ & $\Pi_\phi$  & $\Pi_z$ \\
\hline \hline
$7$ & $-8$  &$\rp{7} \sph{8}$  &$\rp{7} \cph{8}$  &$0$  \\
$7$ & $-7$  &$7\rp{6} z \sph{7}$  &$7\rp{6} z \cph{7}$  &$\rp{7} \sph{7}$  \\
$7$ & $-6$  &$\rho^5 (21z^2 - 2\rho^2) \sph{6}$ &$\frac32 \rho^5 (14z^2 - \rho^2) \cph{6}$ &$7\rho^6z \sph{6}$ \\
$7$ & $-5$  &$\frac74 \rho^4z (20z^2 - 7\rho^2) \sph{5}$ &$\frac{35}{4}\rho^4z (4z^2 - \rho^2) \cph{5}$ &$\frac74 \rho^5 (12z^2 - \rho^2) \sph{5}$ \\
$7$ & $-4$  &$\frac74 \rho^3 (20z^4 - 18\rho^2z^2 + \rho^4) \sph{4}$ &$\frac78 \rho^3 (40z^4 - 24\rho^2z^2 + 3\rho^4) \cph{4}$ &$\frac72 \rho^4z (10z^2 - 3\rho^2) \sph{4}$ \\
$7$ & $-3$  &$\frac{7}{16}\rho^2z (48z^4 - 100\rho^2z^2 + 21\rho^4) \sph{3}$ &$\frac{21}{16}\rho^2z (16z^4 - 20\rho^2z^2 + 3\rho^4) \cph{3}$ &$\frac{7}{16}\rho^3 (80z^4 - 60\rho^2z^2 + 3\rho^4) \sph{3}$ \\
$7$ & $-2$  &$\frac{7}{16}\rho (16z^6 - 80\rho^2z^4 + 45\rho^4z^2 - 2\rho^6) \sph{2}$ &$\frac{7}{32}\rho (32z^6 - 80\rho^2z^4 + 30\rho^4z^2 - \rho^6) \cph{2}$ &$\frac{7}{16}\rho^2z (48z^4 - 80\rho^2z^2 + 15\rho^4) \sph{2}$ \\
$7$ & $-1$  &$\frac{1}{64}z (64z^6 - 1008\rho^2z^4 + 1400\rho^4z^2 - 245\rho^6) \sphi$ &$\frac{1}{64}z (64z^6 - 336\rho^2z^4 + 280\rho^4z^2 -35\rho^6) \cphi$ &$\frac{7}{64}\rho (64z^6 - 240\rho^2z^4 + 120\rho^4z^2 - 5\rho^6) \sphi$ \\
$7$ &  $0$  &$\frac{7}{128}\rho (-64z^6 + 240\rho^2z^4 - 120\rho^4z^2 + 5\rho^6)$ &$0$ &$\frac{1}{16}z (16z^6 - 168\rho^2z^4 + 210\rho^4z^2 - 35\rho^6)$ \\
$7$ &  $1$  &$\frac{1}{64}z (64z^6 - 1008\rho^2z^4 + 1400\rho^4z^2 - 245\rho^6) \cphi$ &$-\frac{1}{64}z (64z^6 - 336\rho^2z^4 + 280\rho^4z^2 -35\rho^6) \sphi$ &$\frac{7}{64}\rho (64z^6 - 240\rho^2z^4 + 120\rho^4z^2 - 5\rho^6) \cphi$ \\
$7$ &  $2$  &$\frac{7}{16}\rho (16z^6 - 80\rho^2z^4 + 45\rho^4z^2 - 2\rho^6) \cph{2}$ &$-\frac{7}{32}\rho (32z^6 - 80\rho^2z^4 + 30\rho^4z^2 - \rho^6) \sph{2}$ &$\frac{7}{16}\rho^2z (48z^4 - 80\rho^2z^2 + 15\rho^4) \cph{2}$ \\
$7$ &  $3$  &$\frac{7}{16}\rho^2z (48z^4 - 100\rho^2z^2 + 21\rho^4) \cph{3}$ &$-\frac{21}{16}\rho^2z (16z^4 - 20\rho^2z^2 + 3\rho^4) \sph{3}$ &$\frac{7}{16}\rho^3 (80z^4 - 60\rho^2z^2 + 3\rho^4) \cph{3}$ \\
$7$ &  $4$  &$\frac74 \rho^3 (20z^4 - 18\rho^2z^2 + \rho^4) \cph{4}$ &$-\frac78 \rho^3 (40z^4 - 24\rho^2z^2 + 3\rho^4) \sph{4}$ &$\frac72 \rho^4z (10z^2 - 3\rho^2) \cph{4}$ \\
$7$ &  $5$  &$\frac74 \rho^4z (20z^2 - 7\rho^2) \cph{5}$ &$-\frac{35}{4}\rho^4z (4z^2 - \rho^2) \sph{5}$ &$\frac74 \rho^5 (12z^2 - \rho^2) \cph{5}$ \\
$7$ &  $6$  &$\rho^5 (21z^2 - 2\rho^2) \cph{6}$ &$-\frac32 \rho^5 (14z^2 - \rho^2) \sph{6}$ &$7\rho^6z \cph{6}$ \\
$7$ &  $7$  &$7\rp{6} z \cph{7}$  &$-7\rp{6} z \sph{7}$  &$\rp{7} \cph{7}$  \\
$7$ &  $8$  &$\rp{7} \cph{8}$  &$-\rp{7} \sph{8}$  &$0$  \\
\hline
\end{tabular}}
\end{table*}

\clearpage
\section{Transverse inhomogeneity\label{App:Bt2Expression}}
In this appendix we give the expression for the averaged squared transverse field inhomogeneity, 
\begin{equation}
\av{B_{\rm T}^2} = \av{(B_x-\av{B_x})^2+(B_y-\av{B_y})^2}, 
\end{equation}
in terms of the generalized gradients $G_{l,m}$ up to order $l=4$ for a cylindrical precession chamber of radius $R$ and height $H$. 
Note that in the analysis, all contributions up to order $l=6$ were considered, having being derived using a computer algebra program, though they are too large to reasonably include here and contribute little to the discussion. 
It can be expressed as a sum of several contributions, one being the contributions of $l$ order modes and the other being the contributions of interferences between modes with different order $l$ and same $\phi$-dependence $m$: 
\begin{equation}
\begin{aligned}
\av{B_{\rm T}^2} =& \av{B_{\rm T}^2}_{\rm 1O} + \av{B_{\rm T}^2}_{\rm 2O} + \av{B_{\rm T}^2}_{\rm 3O} + \av{B_{\rm T}^2}_{\rm 4O} \\
&+ \av{B_T^2}_{\rm 3I1} + \av{B_T^2}_{\rm 4I2}.
\end{aligned}
\end{equation}
The linear-order contribution is: 
\begin{equation}
\begin{aligned}
\av{B_{\rm T}^2}_{\rm 1O} = &\frac{R^2}{2} \left( G_{1,-2}^2+ G_{1,2}^2+\frac14 G_{1,0}^2 \right) \\
+ &\frac{H^2}{12} \left( G_{1,-1}^2+G_{1,1}^2 \right).
\end{aligned}
\end{equation}
The quadratic-order contribution is: 
\begin{equation}
\begin{aligned}
\av{B_{\rm T}^2}_{\rm 2O} 
= & \frac{R^4}{3} \left( G_{2,-3}^2+G_{2,3}^2 \right) \\ 
+ & \frac{R^2 H^2}{12} \left(2G_{2,-2}^2 + 2G_{2,2}^2 + \frac12 G_{2,0}^2 \right) \\
+ & \left( \frac{R^4}{24}+\frac{H^4}{180} \right) \left( G_{2,-1}^2+G_{2,1}^2 \right).
\end{aligned}
\end{equation}
The cubic-order contribution is: 
\begin{equation}
\begin{aligned}
\av{B_{\rm T}^2}_{\rm 3O} 
= & \frac{R^6}{4} \left( G_{3,-4}^2+G_{3,4}^2 \right) \\ 
+ & \frac{R^4 H^2}{4} \left( G_{3,-3}^2+G_{3,3}^2 \right) \\ 
+ & \left( \frac{5R^6}{32} - \frac{R^4H^2}{8} + \frac{9R^2 H^4}{160} \right) \left( G_{3,-2}^2+G_{3,2}^2 \right) \\
+ & \left( \frac{5R^4H^2}{64} - \frac{3R^2H^4}{160} + \frac{H^6}{448} \right) \left( G_{3,-1}^2+G_{3,1}^2 \right) \\
+ & \left( \frac{9R^6}{256} - \frac{R^4H^2}{32} + \frac{9 R^2H^4}{640} \right) G_{3,0}^2.
\end{aligned}
\end{equation}
The fourth order contribution is:
\begin{equation}
\begin{aligned}
\av{B_{\rm T}^2}_{\rm 4O} = \ &\frac{R^8}{5} \Gsq{4}{5} \\
+ \ &\frac{R^6 H^2}{3} \Gsq{4}{4} \\
+ \ &\frac{1}{4} \left( \frac{17 R^8}{20} - R^6 H^2 + \frac{3 R^4 H^4}{5} \right) \Gsq{4}{3} \\
+ \ &\frac{1}{2} \left( \frac{5R^6 H^2}{12} - \frac{R^4 H^4}{5} + \frac{R^2 H^6}{28} \right) \Gsq{4}{2} \\
+ \ &\frac{1}{8} \left( \frac{R^8}{5} - \frac{R^6 H^2}{4} + \frac{R^4 H^4}{4} - \frac{R^2 H^6}{35} + \frac{H^8}{450} \right) \left(
\begin{aligned}
&G_{4,-1}^2 \\
&+G_{4,1}^2
\end{aligned}
\right)\\
+ \ &\frac{1}{8} \left( \frac{3R^6 H^2}{8} - \frac{R^4 H^4}{5} + \frac{R^2 H^6}{28} \right) G_{4,0}^2.
\end{aligned}
\end{equation}

Finally, there are the interference terms, one between the linear and cubic modes and another between quadratic and fourth orders. Note that the odd $l$ modes do not interfere with the even ones. 
\begin{eqnarray}
\av{B_{\rm T}^2}_{\rm 3I1} 
\nonumber
& = & \left(- \frac{R^4}{2}+\frac{R^2H^2}{4} \right) \left( G_{1,-2} G_{3,-2} + G_{1,2} G_{3,2} \right) \\
& & + \left( -\frac{R^2H^2}{8} + \frac{H^4}{40} \right) (G_{1,-1} G_{3,-1} + G_{1,1} G_{3,1}) \nonumber \\
& & + \frac14 \left(- \frac{R^4}{2}+\frac{R^2H^2}{4} \right) \left( G_{1,0} G_{3,0} \right).
\end{eqnarray}
\begin{equation}
\begin{aligned}
\av{B_{\rm T}^2}_{\rm 4I2} = \ &\left( -\frac{R^6}{2} + \frac{R^4 H^2}{3} \right) ( G_{2,-3} G_{4,-3} + G_{2,3} G_{4,3} ) \\
+ \ &\left( -\frac{R^4 H^2}{3} + \frac{R^2 H^4}{10} \right) ( G_{2,-2} G_{4,-2} + G_{2,2} G_{4,2} ) \\
+ \ &\frac{1}{4} \left( -\frac{R^6}{4} + \frac{R^4H^2}{6} - \frac{R^2 H^4}{15} + \frac{H^6}{105} \right) \left( 
\begin{aligned}
&G_{2,-1} G_{4,-1} \\
&+ G_{2,1} G_{4,1} 
\end{aligned}
\right) \\
+ \ &\frac{1}{4} \left( -\frac{R^4 H^2}{3} + \frac{R^2 H^4}{10} \right) G_{2,0} G_{4,0} .
\end{aligned}
\end{equation}

\section{Earth's rotation\label{sec:rotation}}
Though not strictly related to the inhomogeneity of the magnetic field, one effect relevant to the correction strategy arises from the Earth's rotation \cite{Lamoreaux2007}. The neutron EDM measurement took place at the Paul Scherrer Institute in Switzerland. The main $B_0$ magnetic field pointed approximately up or down, as defined by gravity. As such, there was an angle between the Earth's rotational axis and the quantization axis of the system of $\theta = \ang{42.4833}$. Thus, the neutron EDM measurement was effectively taken in a rotating reference frame, effectively shifting the measured neutron and mercury frequencies, and thus $\R$. The correction can be computed as
\begin{equation}
    \delta_{\mathrm{earth}} = \mp \left( \frac{f_{\mathrm{earth}}}{f_{\mathrm{n}}} + \frac{f_{\mathrm{earth}}}{f_{\mathrm{Hg}}} \right) \cos \theta.
\end{equation}
The shift is opposite for each direction of $B_0$. While this does not directly cause a false-EDM like systematic effect as the frequency shift does not depend on the electric field direction, if not considered it can bias the correction strategy described in Subsection \ref{sec:CorrStrat} to produce an error of the order $-2.6 \times 10^{-26} e$~cm.

\end{document}